%% file: space_time_beam.tex
\documentclass[a4paper,fleqn]{cas-sc}

\input{tex/packages.tex}
\input{tex/shortcuts.tex}

\begin{document}
\let\WriteBookmarks\relax
\def\floatpagepagefraction{1}
\def\textpagefraction{.001}
\shorttitle{A space-time finite element formulation for geometrically exact shear-deformable beams}
\shortauthors{I. Steinbrecher and A. Humer}

\title [mode = title]{A space-time finite element formulation for geometrically exact shear-deformable beams}

\author[1]{Ivo Steinbrecher}[orcid=0000-0002-2172-6826]
\ead{ivo.steinbrecher@unibw.de}

\credit{Conceptualization, Methodology, Formal analysis, Software, Validation, Data curation, Investigation, Writing -- original draft, Writing -- review \& editing,     Visualization}

\affiliation[1]{organization={Institute for Mathematics and Computer-Based Simulation (IMCS), University of the Bundeswehr Munich},
                addressline={Werner-Heisenberg-Weg 39},
                postcode={85577},
                postcodesep={},
                city={Neubiberg},
                country={Germany}}

\author[2]{Alexander Humer}[orcid=0000-0001-8510-5423]
\ead{alexander.humer@jku.at}

\affiliation[2]{organization={Institute of Technical Mechanics, Johannes Kepler University Linz},
                addressline={Altenberger Straße 69}, 
                postcode={4040}, 
                postcodesep={}, 
                city={Linz},
                country={Austria}}

\credit{Methodology, Formal analysis, Software, Validation, Data curation, Investigation, Writing -- original draft, Writing -- review \& editing}

\begin{abstract}
We present a space-time finite element formulation for the transient analysis of geometrically exact shear-deformable beams.
The formulation treats space and time within a unified finite element framework and employs a mixed approach in which translational and angular velocities are introduced as independent fields.
The rotation field is parametrized using nodal rotation \mbox{(pseudo-)}vectors and a suitable objective interpolation of the triad field is employed.
The resulting formulation supports generally curved beam structures, including geometric kinks and rigid joints.
Furthermore, time-dependent material domains can be represented directly through the geometry of the space-time surface, enabling the treatment of non-material boundaries without modifying the underlying beam formulation.
Numerical examples demonstrate convergence to solutions obtained with classical time-stepping schemes, objectivity of the formulation, and the influence of the space-time stabilization.
Finally, the sliding spaghetti problem illustrates the applicability of the proposed approach to axially moving continua.
\end{abstract}




\maketitle

\section{Introduction}
Dynamic simulations of beams are essential for the analysis of vibrations and transient response of slender structural components in numerous civil and mechanical engineering applications.
In problems involving large displacements and rotations, geometrically exact beam models are required to accurately capture their nonlinear kinematics.
In the present paper, we develop a space-time formulation for the transient analysis of geometrically exact shear-deformable beams.
Conventionally, approximate solutions to parabolic or hyperbolic problems are constructed in a two-stage approach, which is also known as (vertical) method of lines~\cite{Hairer1993}: A set of ordinary differential equations (in time) is first derived using some favored method for the semi-discretization of the spatial domain. 
Subsequently, a time-stepping method is used to evolve the solution (starting from initial conditions) in time.
Space-time formulations, on the other hand, aim at a unified numerical treatment of the spatial and temporal variables.
In the present work, both domains are discretized simultaneously using the finite element method.
Space and time are thereby treated as coordinates of a unified space-time domain.

Our focus lies on the geometrically exact theory for shear-deformable beams, which is commonly referred to as ``geometrically exact beam theory''.
The theoretical framework is based on seminal contributions by Reissner, Antman and Simo~\cite{Simo1985,Simo1986a, Simo1986b, Simo1988}.
Extending Timoshenko's beam theory~\cite{Timoshenko1921} to problems involving large deformations (hence geometrically exact), Reissner~\cite{Reissner1972} and Antman~\cite{Antman1973} made essential contributions to the theoretical framework for the description of shear-deformable beams. 
Simo and Vu-Quoc subsequently provided finite element formulations for both the three-dimensional~(3D)~\cite{Simo1985,Simo1988} and planar~\cite{Simo1986a,Simo1986b} cases, which continue to influence developments on the numerical analysis of beams to this day.
Building on their foundation, formulations based on absolute rotations, see, e.g.,~\cite{Cardona1988, Ibrahimbegovic1995}, and incremental formulations~\cite{ibrahimbegovic1997a} were proposed in subsequent years.
In~3D problems, the interpolation of rotations is delicate, since it does not adhere to the additive nature of interpolation in linear configuration spaces.
Irrespective of the chosen parametrization of rotations, objectivity of strain measures needs to be preserved by the interpolation scheme as addressed by the seminal works of Crisfield and Jeleni{\'c}~\cite{Crisfield1999, Jelenic1999}.
While rotation vectors are employed for the parametrization of rotations in the present work, alternative formulations based on unit quaternions have been proposed by Zupan et al.~\cite{Zupan2013}, Wasmer and Betsch~\cite{Wasmer2024}, and Herrmann et al.~\cite{Herrmann2026}.
We also mention formulations based on the concept of the special Euclidean group~\cite{Sonneville2017,Harsch2023}, which treat translations and rotations in a unified manner combining them into the notion of motion tensors.
Lately, Humer et al.~\cite{Humer2026} proposed a (partially) mixed formulation for geometrically exact beams that establishes objectivity on the shape-function level through a decomposition of discontinuously interpolated rotations.
The issue of objectivity and path-independence of the deformation was also addressed by Meier et al. in a series of contributions on geometrically exact Kirchhoff-type beams~\cite{Meier2014, Meier2015, Meier2019}, which typically require $C^1$-continuous interpolation functions owing to the constraint of vanishing shear deformation.

Just like finite element formulations for beams, space-time methods are by no means novel either.
In his pioneering contributions, Oden~\cite{Oden1969a, Oden1969b} stated \emph{``that finite
element models of functions of any type of independent variable [...] can be constructed''}, and \emph{``one can develop finite element models in four-dimensional space-time''}.
As a first example to illustrate the concept of a uniform discretization of space and time, he chose the one-dimensional wave equation modeling longitudinal vibrations of an elastic bar.
A comprehensive overview of the history of space-time methods, which have seen a rise in popularity in recent years, is given in the recent paper on space-time approaches for rigid bodies (and systems thereof) by Hesch et al.~\cite{Hesch2023}.
In the context of rigid bodies, one has to deal with both translational and rotational degrees of freedom, which is also the case for the present problem of geometrically exact beams. 
Unlike the proposed space-time formulation for beams, however, rotations were parametrized by means of directors, i.e., a triad of vectors whose orthonormality is realized through Lagrange multipliers, in~\cite{Hesch2023}.
Isogeometric space-time analysis (IGA) was applied to parabolic problems by Langer et al.~\cite{Langer2016}, who introduced time-upwind test functions for stabilization, which we also adopt in the proposed approach.
A Galerkin-based space-time formulation for a variety of problems including fluid dynamics and nonlinear elasticity was proposed by Hesch et al.~\cite{Hesch2017}.
Schuß et al.~\cite{Schuss2023} presented a mixed space-time formulation for nonlinear problems in elasticity using conventional finite elements.
More recently, Xu et al.~\cite{Xu2025} combined space-time methods with the virtual element method.
An additional advantage of space-time methods is their ability to naturally represent time-dependent spatial domains.
Rather than treating the motion of the spatial domain separately, its temporal evolution can be represented directly by the geometry of the space-time domain.
This capability was demonstrated by von Danwitz et al.~\cite{Danwitz2023} for advection-diffusion problems on moving and topology-changing spatial domains.

Space-time methods are particularly attractive for beams, which are represented as material lines parametrized by a single spatial coordinate.
Their evolution in time therefore defines a two-dimensional~(2D) space-time surface, which can be discretized using conventional surface finite elements.
Grohmann et al.~\cite{Grohmann2001} presented both a single-field and a mixed space-time method for planar problems of Timoshenko beams showing a von-Karman type nonlinearity.
Specifically, they developed time-discontinuous Galerkin least-squares approaches for the two formulations, which were augmented by stabilization terms to suppress spurious vibrations.
Moreover, an adaptive refinement of space-time discretization was used for computational efficiency.
A space-time finite element formulation for geometrically exact strings was introduced by Ströhle and Betsch~\cite{Stroehle2022}.
Their work also demonstrates the potential of simultaneous space-time discretizations for inverse dynamics, where prescribed trajectories and the corresponding actuating forces are considered within a unified space-time formulation.
More recently, Hesch and Sahin~\cite{Hesch2026} presented a mixed configuration-momentum-based space-time formulation for the nonlinear dynamics of initially straight, geometrically exact Cosserat beams.
While existing space-time formulations for beams focus on problems with a fixed material domain, the space-time setting also provides the possibility to represent time-dependent material domains.
Such problems arise, for example, for sliding structures and axially moving continua, where the boundaries of the considered spatial domain do not correspond to fixed material points~\cite{Humer2013,Steinbrecher2017,Humer2020}.
To the best of the authors' knowledge, this capability has not yet been explored in the context of space-time formulations for geometrically exact beams.

The present paper develops a fully nonlinear space-time formulation for geometrically exact shear-deformable beams based on a rotation-vector interpolation of the triad field.
The employed rotation interpolation preserves objectivity and permits the treatment of arbitrarily curved beam structures, including geometric kinks and rigid joints.
The proposed formulation adopts a mixed framework in which the translational and angular velocities are introduced as independent fields.
In addition to providing a unified discretization of the spatial and temporal domains, the formulation allows time-dependent material domains to be represented directly through the reference geometry of the space-time surface.
This enables problems involving non-material boundaries to be treated without modifying the underlying beam formulation, as demonstrated for a variant of the sliding spaghetti problem.

The remainder of this paper is organized as follows.
In \Cref{sec:governing_equations}, we review the mathematical treatment of finite rotations and the governing equations of geometrically exact Simo--Reissner beams before deriving the proposed mixed space-time formulation.
\Cref{sec:discretization} presents the finite element discretization, including the treatment of rotations, stabilization, space-time mesh generation, and numerical integration.
The numerical properties and capabilities of the formulation are investigated in \Cref{sec:numerical_examples} using an initially curved beam, a free-floating L-shaped beam, and the sliding spaghetti problem.
Finally, \Cref{sec:conclusion} summarizes the main findings.

\section{Governing equations}
\label{sec:governing_equations}

\subsection{Large rotations}
\label{sec:large_rotations}

This section provides a concise overview of the mathematical treatment of finite rotations.
For a more comprehensive discussion, see, e.g.,~\cite{Simo1986b, Cardona1988, Ibrahimbegovic1995, Romero2004, Meier2019, Betsch1998, Holzinger2021}.
We begin by introducing a rotation tensor (triad)
\begin{equation}
\label{eq:large_rotations:rottensor}
\triad = \matrix{\gtriad{1}, \gtriad{2}, \gtriad{3}} \in \SO,
\end{equation}
where $\SO$ denotes the special orthogonal group.
The columns~$\gtriad{i}$ form an orthonormal triad~$\triad$ that maps the Cartesian basis vectors~$\tns{e}_{i}$ onto~$\gtriad{i}$.
A rotation (pseudo-)vector~$\rotvec$ is employed for parametrization of the triad, \ie,~$\triad = \triad\br[1]{\rotvec}$.
The rotation vector~$\rotvec$ represents a rotation of magnitude~$\rotvecnorm = \norm{\rotvec}$ about the axis oriented along the unit vector~$\rotvecaxis = \rotvec / \norm{\rotvec}$.
Using Rodrigues' formula~\cite{Argyris1982}, the triad can be expressed as
\begin{equation}
\label{eq:large_rotations:rodrigues}
\triad \br[1]{\rotvec}
= \exp \Sskew{\rotvec}
= \tnssI
  + \sin\rotvecnorm \, \Sskew{\rotvecaxis}
  + (1-\cos\rotvecnorm) \, \SskewOp^2\br{\rotvecaxis},
\end{equation}
where $\exp \placeholder$ denotes the matrix exponential.
In the previous equation,~$\SskewOp \in \SOtan$ is an operator such that~$\Sskew{\tns{a}}\,\tns{b} = \tns{a} \times \tns{b}$ for all $\tns{a},\tns{b} \in \R{3}$.
The tangent space of $\SO$ at the identity, denoted by $\SOtan$, is the Lie algebra of skew-symmetric tensors.
The inverse of \eqref{eq:large_rotations:rodrigues}, \ie, the rotation vector as a function of the rotation tensor, is denoted~$\rotvec\br[1]{\triad} = \rv \br[1]{\triad}$.
In practice, the extraction of~$\rotvec$ can be performed using Spurrier's algorithm~\cite{Spurrier1978}.

Two rotation tensors~$\triad_1 = \triad\br[1]{\rotvec_{1}}$ and~$\triad_2 = \triad\br[1]{\rotvec_{2}}$ with their respective rotation vectors~$\rotvec_{1}$ and~$\rotvec_{2}$ can be related by a relative rotation~$\triad_{21} = \triad\br[1]{\rotvec_{21}}$ according to
\begin{equation}
\label{left_multi}
\triad_{2} = \triad_{21} \triad_1
\quad \Leftrightarrow \quad
\triad_{21} = \triad_{2} \triad_1\tr,
\end{equation}
with~$\triad\tr = \triad\inv$ being the defining property of all~$\triad \in \SO$.
Hence, the (non-additive) rotation vector~$\rotvec_{21} = \rv(\triad_{21}) \ne \rotvec_{2} - \rotvec_{1}$ describes the relative rotation between~$\triad_1$ and~$\triad_2$.

Next, we consider infinitesimal variations of the rotation tensor.
These can be expressed either by an additive infinitesimal variation~$\drotvec$ of the rotation vector,
\begin{equation}
\label{eq:add_variation}
\delta \triad
= \at{\ptfrac{}{\epsilon}}_{\epsilon=0} \triad\br[1]{\rotvec + \epsilon \drotvec}
= \pfrac{\triad\br[1]{\rotvec}}{\rotvec} \drotvec,
\end{equation}
or equivalently, by a multiplicative infinitesimal rotation (the spin vector)~$\drotmult$,
\begin{equation}
\label{eq:mult_variation}
\delta \triad
= \at{\ptfrac{}{\epsilon}}_{\epsilon=0} \triad\br[1]{\epsilon \drotmult} \triad\br[1]{\rotvec}
= \Sskew{\drotmult} \triad\br[1]{\rotvec}.
\end{equation}
From~\eqref{eq:mult_variation}, the variations of the triad basis vectors follow as
\begin{equation}
\dgtriad{i} = \drotmult \times \gtriad{i}.
\end{equation}
The infinitesimal additive and multiplicative variations are related through
\begin{equation}
\label{eq:large_rotations:rot_mult_variation}
\drotvec = \Ttrans \br[1]{\rotvec} \drotmult,
\end{equation}
where~$\Ttrans(\rotvec)$ denotes the transformation matrix~\cite{Simo1988},
\begin{equation}
\label{eq:large_rotations:transformation_matrix}
\Ttrans(\rotvec) =
\frac{1}{\rotvecnorm^2} \rotvec \rotvec\tr
- \frac{1}{2} \Sskew{\rotvec}
+ \frac{\rotvecnorm}{2 \tan(\rotvecnorm/2)}
  \left( \tnssI - \frac{1}{\rotvecnorm^2} \rotvec \rotvec\tr \right).
\end{equation}

\subsection{\Gexact beam formulation}
\label{sec:beam_theory}

In what follows, we outline the equations that govern the spatial motion of beams within the framework of the Simo--Reissner theory for shear-deformable beams.
We consider a beam as a one-dimensional continuum that is parametrized by the arc-length coordinate $\sbeam$ of its centerline in the initial (undeformed) configuration.
The beam is embedded in the three-dimensional~(3D) Euclidean space, where it describes a spatial curve of initial length $\beamlength$. 
The position of the centerline in the initial configuration is denoted by
\begin{equation}
    \label{eq:ref_config}
    \sbeam \mapsto \rbeamO \in \R{3} , \qquad \sbeam \in \brr{0, \beamlength} =: \domainArcLength \subset \R{} ,
\end{equation}
where the subscript `0' indicates quantities that refer to the undeformed state.
The cross sections of the beam can be regarded as rigid bodies attached to its centerline.
Cross sections are spanned by unit vectors $\gtriadO{2}$ and $\gtriadO{3}$, which together with the unit normal tangential to the centerline $\gtriadO{1} = \dds{\rbeamO}$, form the triad
\begin{equation}
    \sbeam \mapsto \triadO = \gtriadO{i} \otimes \ei = \brr[1]{\gtriadO{1}, \gtriadO{2}, \gtriadO{3}} \in \SO ,
\end{equation}
which describes the rotation from the global Cartesian frame $\{ \ex, \ey, \ez \}$.

The beam's deformed configuration is described by the instantaneous position (at time $\timeVar$) of its centerline $s, t \mapsto \rbeam \in \R{3}$ and the orientation of its cross sections $s, t \mapsto \triad \in \SO$, which, according to Timoshenko's hypothesis, remain plane but not necessarily perpendicular to the centerline.

Clearly, the temporal evolution of deformation lies in the focus of interest as it comes to space-time methods.
For this purpose, let $\domainTime := \brr{0,\timeEnd} \subset \R{}$ denote the time-domain of interest, i.e.,~$t \in \domainTime$. 
The velocity field $s, t \mapsto \vBeam$ represents the (temporal) rate of change of the position of the beam's centerline; the spatial angular velocity vector $\angularVelocity$
is implicitly defined in terms of the Lie algebra as
\begin{align} \label{eq:angular_velocity}
    \vBeam &= \ddt{\rbeam} , &
    \Sskew{\angularVelocity} &= \dot{\triad} \triad\tr ,
\end{align}
where superimposed dots indicate material time derivatives.
The material angular velocity $\angularVelocityMat$ is defined in terms of its spatial counterpart as
\begin{align} \label{eq:angular_velocity_mat}
    \angularVelocityMat &= \triad\tr \angularVelocity , &
    \Sskew{\angularVelocityMat} &= \triad\tr \dot{\triad} .
\end{align}
A cross section's vectors of linear momentum and angular momentum are given by
\begin{align}
    \beamLinearMomentum &= \density A \vBeam , &
    \beamAngularMomentum &= \densityRotation \angularVelocity = \triad \densityRotationMat \angularVelocityMat ,
\end{align} 
where $\density A$ represents the mass per unit length. 
In case of homogeneous cross sections, it is given by the product of (mass) density $\rho$ and area $A$; otherwise, $\rho A$ is to be understood as an effective property.
In the above equation, $\densityRotation$, $\densityRotationMat$ denote the tensor of moment of inertia and its material counterpart, which, when represented in their respective frames, have constant components
\begin{align}
    \densityRotation 
    &= \density I_p \gtriad{1} \dyad \gtriad{1} + \density I_2 \gtriad{2} \dyad \gtriad{2} + \density I_3 \gtriad{3} \dyad \gtriad{3} , &
    \densityRotationMat &= \triad\tr \densityRotation \triad ,
\end{align}
where $I_p$ is the polar area moment about the centerline and $I_2$, $I_3$ are second area moments about the principal axes of the cross-section frame.
The rates of the momenta define the inertia forces and moments
\begin{align}
    \beamForceIntertia &=  - \ddt{\beamLinearMomentum} = - \density \crossSectionArea \dddt{\rbeam} , &
    \beamMomentIntertia &= - \ddt{\beamAngularMomentum} = - \triad \left(\Sskew{ \angularVelocityMat} \densityRotationMat \angularVelocityMat + \densityRotationMat \ddt{\angularVelocityMat} \right) .
\end{align}

The strong form of the governing equations is given by the local balance of laws of linear momentum and angular momentum.
Let $\beamForce$ and $\beamMoment$ denote the vectors of resultant forces and moments with respect to the beam's centerline $\rbeam$. 
The local balance laws state
\begin{align} \label{eq:balance_momentum}
    \dds{\beamForce} + \beamBodyLoad + \beamForceIntertia &= \tnsO , &
    \dds{\beamMoment} + \dds{\rbeam} \times \beamForce + \beamBodyMoment + \beamMomentIntertia &= \tnsO ,
\end{align}
where primes $\dds{\placeholder} = \pfracinline{\placeholder}{\sbeam}$ indicate derivatives with respect to the (material) arc-length coordinate $\sbeam$ in the reference configuration.
In the above equations, $\beamBodyLoad$ and $\beamBodyMoment$ are externally imposed forces and moments per unit (undeformed) length.

At the ends of the beam $\domainBoundary = \partial \domainArcLength$, concentrated forces $\beamForceExt$ and moments $\beamMomentExt$ are applied, which constitute natural boundary conditions to the local balance laws,
\begin{align}
    \beamForce(0) &= {\beamForce}_{\sigma, 0} , & 
    \beamForce(L) &= {\beamForce}_{\sigma, L} , &
    \beamMoment(0) &= {\beamMoment}_{\sigma, 0} , & 
    \beamMoment(L) &= {\beamMoment}_{\sigma, L} .
\end{align}

To close the set of equations, we need to specify the constitutive relations between the internal forces and moments and the beam's deformation. 
For this purpose, we first introduce the spatial curvature $\beamCurvature$ as the spatial rate of change of the cross section's orientation along the beam's axis,
\begin{align} \label{eq:beamCurvature}
    \Sskew{\beamCurvature} &= \dds{\triad} \triad\tr .
\end{align}
In analogy to the angular velocity~\eqref{eq:angular_velocity_mat}, the material curvature $\beamCurvatureMat$ follows upon rotating $\beamCurvature$ back to the global frame, 
\begin{align} \label{eq:beamCurvatureMat}
    \beamCurvatureMat &= \triad\tr \beamCurvature , &
    \Sskew{\beamCurvatureMat} &= \triad\tr \dds{\triad} .
\end{align}
In the Simo--Reissner beam theory, the generalized strain $\srBendingMat$ related to the twist and bending of the beam's axis is defined as the difference of material curvatures in the current configuration and the reference configuration, respectively,
\begin{align}
    \srBendingMat &= \beamCurvatureMat - \beamCurvatureMat_0 , &
    \Sskew{\beamCurvatureMat_0} = \triadO\tr \dds{\triadO} .
\end{align}
Assuming the beam to be unstretched in the reference configuration, the generalized force strain $\srTensionMat$, on the other hand, is given by
\begin{equation} \label{eq:force_strain}
    \srTensionMat = \triad\tr \dds{\rbeam} - \ex .
\end{equation}
The corresponding spatial strain measures then follow upon a push-forward of the material strains by means of $\triad$ as
\begin{align}
    \srBending &= \triad \srBendingMat = \beamCurvature - \triad \triad_0\tr \beamCurvature_0 , &
    \srTension &= \triad \srTensionMat = \dds{\rbeam} - \gtriad{1}.
\end{align}
In what follows, we assume the simplest possible constitutive relation, i.e., a linear relation between the cross-sectional stress resultants and their conjugate strain measures. 
The material response is then governed by the following relations for the resultant moments and forces, respectively,
\begin{align}
    \beamMoment &= \srBendingC \srBending = \triad \srBendingCMat \srBendingMat , &
    \beamForce &= \srTensionC \srTension = \triad \srTensionCMat \srTensionMat ,
\end{align}
with the second-order tensor of torsional and bending stiffnesses in spatial and material form given by
\begin{align}
    \srBendingC &= GI_T \gtriad{1} \otimes \gtriad{1} + EI_2 \gtriad{2} \otimes \gtriad{2} + EI_3 \gtriad{3} \otimes \gtriad{3} , &
    \srBendingCMat &= \triad\tr \srBendingC \triad ; 
\end{align}
for extensional and shear stiffnesses, we assume
\begin{align}
    \srTensionC &= EA \gtriad{1} \otimes \gtriad{1} + k_2GA \gtriad{2} \otimes \gtriad{2} + k_3GA \gtriad{3} \otimes \gtriad{3} , &
    \srTensionCMat &= \triad\tr \srTensionC \triad ,
\end{align}
where, for homogeneous cross sections, stiffnesses are related to Young's modulus $E$, the shear modulus $G$, the torsional moment of inertia $I_T$ and shear correction factors $k_2$, $k_3$.

\subsection{Space-time formulation}
\label{sec:ST_formulation}

The local balance equations~\eqref{eq:balance_momentum} of momentum and angular momentum serve as starting point for the derivation of the space-time formulation.
The weak form of the balance laws---as underlying conventional finite element discretizations in the spatial domain---is formally obtained by projecting the balance laws onto fields of admissible displacements $\drbeam$ and rotations $\drotmult$, respectively, followed by integration over the spatial domain of the body:
\begin{align}
    \intSpace{\drbeam\tr \left( \dds{\beamForce} + \beamBodyLoad + \beamForceIntertia \right) + \drotmult\tr \left( \dds{\beamMoment} + \dds{\rbeam} \times \beamForce + \beamBodyMoment + \beamMomentIntertia \right)} = 0 .
\end{align}
Admissibility requires displacements $\drbeam$ and rotations $\drotmult$ to be compatible with kinematic boundary conditions on displacements and rotations.
Upon integration by parts in the stress resultants, we arrive at what, from a mechanical perspective, constitutes the principle of virtual work. 
It states that the sum of the virtual work performed by internal forces $\dWint$, the work of external forces $\dWext$, and the work of inertia forces $\dWInertia$ vanishes for any kinematically admissible displacement from the current configuration, i.e.,
\begin{align} \label{eq:pvw}
    \dWint + \dWext + \dWInertia = 0 .
\end{align}
where the virtual work of the internal forces is given by
\begin{align}
    \dWint
        &= - \intSpace{
        {\dds{\drbeam}}\tr \beamForce 
        - \drotmult\tr \br{\dds{\rbeam} \times \beamForce}
        + {\dds{\drotmult}}\tr \beamMoment 
        }
        .
\end{align}
The virtual work of the external forces is composed of distributed loads and concentrated forces acting at the beam's tips,
\begin{align}
    \dWext = \intSpace{
        \drbeam\tr \beamBodyLoad 
        + \drotmult\tr \beamBodyMoment 
    }
    + \brr[2]{\drotmult\tr \beamMomentExt + \drbeam\tr \beamForceExt}_{0}^{L} ,
\end{align} 
and the virtual work of the inertia forces is given by
\begin{align}
    \label{eq:pvw_inertia}
    \dWInertia = \intSpace{
        \drbeam\tr \beamForceIntertia 
        + \drotmult\tr \beamMomentIntertia
    } .
\end{align}

Up to this point, we have assumed configurations to be fixed with respect to the material body.
Then, the material points contained in the spatial domain of interest~$\domainArcLength$ remain the same throughout the course of motion, cf.~\eqref{eq:ref_config}.
While this is true for many problems in structural mechanics, the fact that the (spatial) computational domain is not fixed to the material points is a distinguishing feature of problems of sliding structures and axially moving continua~\cite{Humer2020}, or problems of moving loads.
Consider, for instance, the well-known sliding spaghetti problem~\cite{Carrier1949}, which is inspired by a thin, long piece of pasta being sucked in.
The material length of the part yet to be consumed decreases over time.
In the numerical modeling, boundary conditions need to be prescribed at varying material points, which poses challenges in conventional mesh-based approaches, see, e.g.,~\cite{Steinbrecher2017}.
Typically, some sort of mapping between the fixed computational domain and the instantaneous, varying material domain is employed, at the cost of complex inertia terms~\cite{Humer2013,Humer2020,Vetyukov2018}.

The space-time setting developed in what follows allows us to deal with problems involving varying material domains in a straightforward way.
Assuming the motion of the material points at the boundaries of the spatial domain to be prescribed kinematically, such varying material domains can be represented directly through the geometry of the reference space-time surface, as will be shown by means of the sliding spaghetti problem in \Cref{sec:examples:spaghetti}.
To also include problems of this kind, we re-define the (spatial) domain of the beam centerline as time-dependent, i.e.,~$\domainArcLength(t) := [\sbeama, \sbeamb]$, accepting the slight abuse of notation.
Material points at the boundaries of the spatial domain are no longer fixed, but may be functions in time~$\sbeama$ and~$\sbeamb$.
It is important to distinguish this time dependence of the considered material domain from the reference configuration itself.
The reference position~$\rbeamO(\sbeam)$ of a material point is assumed to be independent of time; only the subset~$\domainArcLength(t)$ of material points represented by the spatial domain may vary over time.

The current configuration~$\domainSpaceTime$ and the reference configuration~$\domainSpaceTimeO$ of the beam centerline define two-dimensional~(2D) manifolds parametrized by the arc-length coordinate~$\sbeam \in \domainArcLength(t)$ and time~$t \in \domainTime$ and embedded in four-dimensional~(4D) space-time.
These manifolds are referred to as the current and reference space-time surfaces, respectively:
\begin{align}
    \domainSpaceTime := \brrr{ \br[1]{\rbeam(s,t), t} \,\middle|\, \sbeam \in \domainArcLength(t),\ t \in \domainTime } \subset \mathbb{R}^4 , & &
    \domainSpaceTimeO := \brrr{ \br[1]{\rbeamO(s), t} \,\middle|\, \sbeam \in \domainArcLength(t),\ t \in \domainTime } \subset \mathbb{R}^4 .
\end{align}
The space-time weak form is obtained analogously to the conventional weak formulation introduced above, with the essential difference that the projected balance laws are integrated over both the spatial and temporal domains.
Prior to integration by parts, this yields
\begin{align}
\dAct
= \intSpaceTime{\drbeam\tr \left( \dds{\beamForce} + \beamBodyLoad + \beamForceIntertia \right) + \drotmult\tr \left( \dds{\beamMoment} + \dds{\rbeam} \times \beamForce + \beamBodyMoment + \beamMomentIntertia \right)} = 0
.
\end{align}
As in the conventional formulation, integration by parts is performed with respect to the spatial coordinate~$\sbeam$ for the terms involving the stress resultants.
No integration by parts is performed in the temporal direction.
The resulting space-time weak form can therefore be expressed in terms of the virtual work contributions introduced in~(\ref{eq:pvw}--\ref{eq:pvw_inertia}) as
\begin{align}
\dAct
=&
\intTime{
    \dWint + \dWext + \dWInertia 
}
\nonumber
\\
=&
- \intSpaceTime{
        {\dds{\drbeam}}\tr \beamForce 
        - \drotmult\tr \br{\dds{\rbeam} \times \beamForce}
        + {\dds{\drotmult}}\tr \beamMoment 
        }
\nonumber
\\
&+ \intSpaceTime{
        \drbeam\tr \beamBodyLoad 
        + \drotmult\tr \beamBodyMoment 
    }
    + \intTime{
        \brr[2]{\drotmult\tr \beamMomentExt + \drbeam\tr \beamForceExt}_{\sbeama}^{\sbeamb}
    }
\nonumber
\\
&+
\intSpaceTime{
        \drbeam\tr \beamForceIntertia 
        + \drotmult\tr \beamMomentIntertia
    }
= 0
.
\end{align}

In the space-time formulation, (spatial) boundary conditions and initial conditions are formulated at the boundaries of $\domainSpaceTimeO$, which we denote by 
\begin{align}
    \boundarySpaceA &:= \brrr{  \br[1]{\rbeamO(\sbeama), t} \,\middle|\, t \in \domainTime }, & 
    \boundarySpaceB &:= \brrr{ \br[1]{\rbeamO(\sbeamb), t} \,\middle|\, t \in \domainTime },
    \\
    \boundaryTimeO &:= \brrr{ \br[1]{\rbeamO(s), 0} \,\middle|\, s \in \domainArcLength(0) }, & 
    \boundaryTimeT &:= \brrr{ \br[1]{\rbeamO(s), \timeEnd} \,\middle|\, s \in \domainArcLength(\timeEnd) }.
\end{align}
The spatial conditions applied at~$\boundarySpaceA$ and~$\boundarySpaceB$ directly carry over from the conventional formulation~\eqref{eq:pvw}.
In addition, the space-time surface contains the temporal boundaries~$\boundaryTimeO$ and~$\boundaryTimeT$, corresponding to the initial and final time, respectively.
At the initial boundary~$\boundaryTimeO$, we prescribe essential conditions on the positions~$\rbeam(s,0) = \bar{\rbeam}(s)$ and velocities~$\dot{\rbeam}(s,0) = \bar{\vBeam}(s)$, as well as on the rotations~$\triad(s,0) = \bar{\triad}(s)$ and angular velocities~$\dot{\triad}(s,0) = \Sskew{\bar{\angularVelocity}(s)}\triad(s,0)$.
No terminal conditions are prescribed at the final boundary~$\boundaryTimeT$.
Here, the superscript~$\bar{\placeholder}$ denotes a prescribed boundary value.

In a formulation based solely on the positions~$\rbeam$ and orientations~$\triad$ as primary fields, the inertia terms involve second-order derivatives with respect to time, which would require~$C^1$-continuous approximations in the temporal direction.
Moreover, initial conditions on the translational and angular velocities involve temporal derivatives of the primary variables, whose direct imposition is not straightforward with standard~$C^0$ finite element approximations.
For the rotational field, this is further complicated by the nonlinear relation between the chosen rotation parametrization and the angular velocity~$\angularVelocity$.
For these reasons, we employ a mixed formulation in which the velocity field~$\vBeam$ and the material angular velocity~$\angularVelocityMat$ serve as independent fields, following the general approach of mixed space-time formulations, cf.~\cite{Schuss2023,Hesch2026,Xu2025,Hesch2023}.
This reduces the highest temporal derivative appearing in the formulation to first order and allows initial conditions on the translational and angular velocities to be imposed directly.
To distinguish the independent velocity fields from those obtained from the configuration fields, we introduce
\begin{align}
    \vBeamFromPos &= \ddt\rbeam ,
    &
    \Sskew{\angularVelocityMatFromRot} &= \triad\tr \ddt\triad .
\end{align}
The corresponding kinematic relations~$\vBeamFromPos = \vBeam$ and~$\angularVelocityMatFromRot = \angularVelocityMat$ are enforced in a weak sense.
The resulting mixed space-time weak form reads
\begin{align}
    \label{eq:space_time_weak_form}
    \dActMixed
    =-&
    \intSpaceTime{
        {\dds{\drbeam}}\tr \beamForce 
        - \drotmult\tr \br{\dds{\rbeam} \times \beamForce}
        + {\dds{\drotmult}}\tr \beamMoment 
    } \nonumber
    \\
    +&\intSpaceTime{
        \drbeam\tr \br{\beamBodyLoad - \rho \crossSectionArea \ddt\vBeam}
        + \drotmult\tr \brrr{\beamBodyMoment - \triad \left(\Sskew{ \angularVelocityMat} \densityRotationMat \angularVelocityMat + \densityRotationMat \ddt{\angularVelocityMat} \right)}
    } \nonumber
    \\
    +&\intTime{
        \brr[2]{\drotmult\tr \beamMomentExt + \drbeam\tr \beamForceExt}_{\sbeama}^{\sbeamb}
    } \nonumber
    \\
    -&\intSpaceTime{
        \density \crossSectionArea \dvBeam\tr \br{\vBeamFromPos - \vBeam}
        +
        \dangularVelocityMat\tr \densityRotationMat \br{\angularVelocityMatFromRot - \angularVelocityMat}
    }
    = 0
    .
\end{align}
The last integral in the equation above weakly enforces the kinematic relations between the configuration fields~$\rbeam$ and~$\triad$ and the independent velocity fields~$\vBeam$ and~$\angularVelocityMat$.
In the proposed space-time formulation, the admissible trial and test functions must satisfy all essential conditions prescribed on the boundary of the space-time domain, including both spatial boundary conditions and initial conditions.
Accordingly, in addition to~$\drbeam$ and~$\drotmult$, the variations of the independent velocity fields~$\dvBeam$ and~$\dangularVelocityMat$ must satisfy the corresponding homogeneous essential conditions.
\begin{remark}
    Note that the usage of letters $\Act$ and $\mathcal S$ for the space-time formulation and its mixed variant is not coincidental. 
    As a matter of fact, the \emph{action} integral of Hamilton's principle is, from a mechanics point of view, a natural starting point for developing a space-time form.
    For the sake of brevity, however, we refer to Schuß et al.~\cite{Schuss2023} for a discussion.
\end{remark}

\subsubsection{Space-time kinematics}

The space-time formulation~\eqref{eq:space_time_weak_form} requires the evaluation of arc-length and time derivatives of field quantities defined on the space-time surface.
The parametrization of the reference space-time surface is, in general, expressed in terms of the convective curvilinear coordinates~$\xisurface$ and~$\etasurface$ rather than directly in terms of the physical coordinates~$\sbeam$ and~$t$.
In the context of the finite element discretization, these curvilinear coordinates correspond to the parameter coordinates of the space-time element.
The purpose of this section is to establish the transformation between these coordinate systems, allowing derivatives of a general field function~$f(\xisurface,\etasurface)$ with respect to~$\sbeam$ and~$t$ to be evaluated.

To simplify the derivations, we define the lifted~4D reference position vector~$\rbeamSTO \in \R{4}$ associated with the spatial reference position~$\rbeamO \in \R{3}$ and time~$t \in \R{}$ as
\begin{equation}
\label{eq:space_time_lifting}
\rbeamSTO = \br{\rbeamO \cdot \ex} \eSTx + \br{\rbeamO \cdot \ey} \eSTy + \br{\rbeamO \cdot \ez} \eSTz + t \eSTt
,
\end{equation}
where~$\eSTx$, $\eSTy$, $\eSTz$, and~$\eSTt$ denote the corresponding orthonormal basis vectors in~$\R{4}$.
This construction embeds the spatial reference position~$\rbeamO$ into the space-time setting by preserving its spatial components and augmenting them with the temporal coordinate~$t$ along~$\eSTt$.

\begin{figure}
  \centering
  \includegraphics[scale=1]{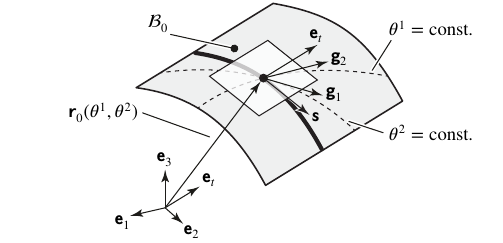}
  \caption{Parametrization of the reference space-time surface~$\domainSpaceTimeO$ by the curvilinear coordinates~$\xisurface$ and~$\etasurface$.
  The thick line indicates the beam configuration containing~$\rbeamSTO(\xisurface,\etasurface)$.}
  \label{fig:discretization_space_time_kinematics}
\end{figure}

The lifted reference position is parametrized by~$\xisurface$ and~$\etasurface$, i.e.,~$\rbeamSTO = \rbeamSTO\br[1]{\xisurface,\etasurface}$, cf.~\Cref{fig:discretization_space_time_kinematics}.
In the following, Einstein's summation convention is employed, where repeated Greek indices imply summation over~$\brrr{1,2}$.
The covariant basis vectors of the space-time surface are defined by
\begin{align}
\tnsCo{\tnsST{g}}{\indexSurface} = \pfrac{\rbeamSTO}{\xisurfaceEinstein}
,
\end{align}
and allow us to compute the covariant metric coefficients of the space-time surface as
\begin{align}
g_{\indexSurface\indexSurfaceAlt} = \tnsCo{\tnsST{g}}{\indexSurface} \cdot \tnsCo{\tnsST{g}}{\indexSurfaceAlt}
.
\end{align}
In terms of these, the contravariant basis vectors are obtained as
\begin{align}
\tnsCont{\tnsST{g}}{\indexSurface} = \tnsCont{g}{\indexSurface\indexSurfaceAlt} \tnsCo{\tnsST{g}}{\indexSurfaceAlt}
\quad
\text{with}
\quad
\tnsBr{g^{\indexSurface\indexSurfaceAlt}} = \tnsBr{g_{\indexSurface\indexSurfaceAlt}}\inv
,
\end{align}
where~$\tnsCont{g}{\indexSurface\indexSurfaceAlt}$ are the contravariant metric coefficients.
Furthermore, the relation between the infinitesimal area elements~$\mathrm d\sbeam\,\mathrm dt$ and~$\mathrm d\xisurface\mathrm d\etasurface$ is given by the Jacobian~$J$:
\begin{align}
\label{eq:kinematic_jacobian}
\mathrm d\sbeam\,\mathrm dt
=
J \mathrm d\xisurface\mathrm d\etasurface
\qquad
\text{with}
\qquad
J
=
\sqrt{
\det\br{\brrr{g_{\indexSurface\indexSurfaceAlt}}}
}
.
\end{align}

Since the reference space-time surface is straight in the temporal direction, the temporal unit vector~$\eSTt$ belongs to its tangent plane.
To complement this direction, we introduce the spatial unit tangent~$\directionSpace$, which lies within the tangent plane and is orthogonal to~$\eSTt$.
The vectors~$\{\directionSpace,\eSTt\}$ thus form an orthonormal basis of the tangent plane.
In other words, moving along~$\eSTt$ corresponds to a pure change in time at a fixed material position, whereas moving along~$\directionSpace$ corresponds to a pure change in arc-length at a fixed time.
The vector~$\directionSpace$ can be constructed as a linear combination of the covariant basis vectors~$\tnsCo{\tnsST{g}}{1}$ and~$\tnsCo{\tnsST{g}}{2}$ such that its projection onto~$\eSTt$ vanishes.
After normalization, this yields
\begin{align}
\directionSpace =
\normalize{
\tnsCo{\tnsST{g}}{1} (\tnsCo{\tnsST{g}}{2} \cdot \eSTt)
-
\tnsCo{\tnsST{g}}{2} (\tnsCo{\tnsST{g}}{1} \cdot \eSTt)
}
.
\end{align}
Consequently, differentiation with respect to arc-length and time can be interpreted as directional derivatives along~$\directionSpace$ and~$\eSTt$, respectively, i.e.,
\begin{align}
\label{eq:derivative_field_function}
\dds{f} =
\pfrac{f(\xisurface, \etasurface)}{\sbeam}
=
\nabla_{\directionSpace} f(\xisurface, \etasurface)
=
\pfrac{f}{\xisurfaceEinstein} \br{\tnsCont{\tnsST{g}}{\indexSurface} \cdot \directionSpace}
\quad
\text{and}
\quad
\ddt{f} =
\pfrac{f(\xisurface, \etasurface)}{t}
=
\nabla_{\eSTt} f(\xisurface, \etasurface)
=
\pfrac{f}{\xisurfaceEinstein} \br{\tnsCont{\tnsST{g}}{\indexSurface} \cdot \eSTt}
.
\end{align}

\begin{remark}
    The space-time lifting~\eqref{eq:space_time_lifting} combines spatial and temporal quantities, which have different physical dimensions.
    For dimensional consistency, the temporal coordinate could be scaled by a constant, cf.~\cite{Hesch2026}.
    It can be shown, that such a scaling factor cancels out in the final equation for the Jacobian~\eqref{eq:kinematic_jacobian} and the directional derivatives~\eqref{eq:derivative_field_function}. 
    For simplicity of presentation, such a scaling is omitted here.
\end{remark}

\section{Discretization}
\label{sec:discretization}

In this section, we present the spatial and temporal discretization of the space-time beam formulation introduced in~\eqref{eq:space_time_weak_form}, which is entirely based on the finite element method.
The continuous space-time domain~$\domainSpaceTimeO$ is discretized into~$\nbeame$ space-time finite elements with a total of~$\nbeamn$ nodes, resulting in the discrete space-time domain~$\domainSpaceTimeOh$.
Topologically, the space-time finite elements are surface elements, where each element~$\ebeam$ connects~$\nbeamne$ nodes.
In this work, we consider four-noded first-order elements (\quadFE{4}) and nine-noded second-order elements (\quadFE{9}).
For the considered \cosserat continuum, two fundamentally different classes of field quantities require interpolation:~(i) field functions which represent elements of a vector space (i.e.,~$\R{}$ and~$\R{3}$) and~(ii) field functions which represent orientations, i.e., elements of~$\SO$.
The first case~(i) is the standard case for most finite element formulations and will be presented in \Cref{sec:discretization_vectorspace}.
The interpolation of the orientation field~(ii) requires special treatment due to the nonlinear structure of the rotation group~$\SO$ and is discussed in \Cref{sec:discretization_orientations}.
The subsequent subsections also describe the stabilization strategy (\Cref{sec:discretization_stabilization}), the assembly of the resulting global system of equations (\Cref{sec:discretization_global_system}), the generation of the space-time mesh (\Cref{sec:discretization_mesh_generation}), and the numerical integration~(\Cref{sec:discretization_integration}).

Throughout this section, a hat~$\hat{\placeholder}$ denotes nodal primary variables, a subscript~$\placeholder_h$ the finite element approximation of a quantity, and a superscript~$\placeholder^{(e)}$ an element-specific quantity.
For example,~$\rbeamn$ denotes the position of a node identified by its global index~$\indexbeam$, whereas~$\rbeamne$ denotes the position of the same node identified by its local index~$\indexbeam$ within element~$\ebeam$.

\subsection{Discretization of vector space quantities}
\label{sec:discretization_vectorspace}

The discretized space-time domain~$\domainSpaceTimeOh$ consists of an interpolation of time and the reference beam centerline position, given by
\begin{align}
t^\ebeam\xisolidDependency &\approx \timehe\xisolidDependency = \sumbeamne{\Nbeamn\xisolidDependency \timene},
\\
\rbeamO^\ebeam\xisolidDependency &\approx \rbeamOhe\xisolidDependency = \sumbeamne{\Nbeamn\xisolidDependency \rbeamOne}.
\end{align}
Here,~$\timene$ and $\rbeamOne$ represent the nodal time and reference position of the~$\indexbeam$-th node of element~$\ebeam$, respectively.
The shape functions~$\Nbeamn$ are Lagrangian finite element shape functions defined over the 2D reference parameter domain~$(\xisolid, \etasolid)$, which parametrizes each space-time surface element.

For the primal field functions that represent objects from a vector space, i.e., beam centerline position and velocity, as well as the cross-section material angular velocity, we employ the same finite element approximation as for the discretization of the space-time domain, i.e.,
\begin{align}
\q^\ebeam\xisolidDependency &\approx \qhe\xisolidDependency
= \sumbeamne{\Nbeamn\xisolidDependency \qne} ,
\end{align}
where~$\q$ represents any of the element fields  $\brrr{\rbeam, \vBeam, \angularVelocityMat}$.
Employing a standard isoparametric finite element approach, the same interpolation is used for the variations of the primal fields:
\begin{align}
\dqq^\ebeam\xisolidDependency &\approx \dqhe\xisolidDependency
= \sumbeamne{\Nbeamn\xisolidDependency \dqne}
,
\end{align}
where~$\dqq$ represents any of the element variations  $\brrr{\drbeam, \dvBeam, \dangularVelocityMat}$.

\subsection{Discretization of rotations}
\label{sec:discretization_orientations}

In this section, we present the interpolation of rotations on the space-time surface.
This is related to an approach originally proposed in~\cite{Shoemake1985} for computer graphics applications and later adapted to finite element formulations of geometrically exact beams in~\cite{Jelenic1999,Crisfield1999}.
The orientation field is parametrized by total rotation (pseudo-)vectors~$\rotvecn$ defined at the nodes of the finite element mesh.
The interpolation is performed in terms of local rotation vectors, which requires the definition of an intermediate triad~$\triadRefe$ for each finite element~$\ebeam$.
Following~\cite{Jelenic1999,Crisfield1999,Meier2019}, this auxiliary triad is commonly referred to as the \emph{reference triad}.
In the present work, however, we use the term \emph{intermediate triad} to avoid confusion with the reference configuration.
The intermediate triad is not unique, however, in order to obtain an objective rotation interpolation, it must itself be defined in an objective manner.
A detailed discussion of suitable choices for the intermediate triad is given in \Cref{sec:discretization_intermediate_triad}.

In contrast to the translational degrees of freedom, rotational degrees of freedom require special treatment whenever multiple beam segments with different reference orientations meet at a common node and share the same rotational degrees of freedom, e.g., at rigid joints or geometric kinks.
The nodal rotation vector~$\rotvecn$ parametrizes a common nodal orientation, from which the orientation of each adjoining beam element is obtained by means of a constant multiplicative rotation offset.
To this end, each element stores a constant rotational offset~$\triadOffsete$ for each of its nodes~$\indexbeam$, relating the shared nodal orientation~$\triad\br[1]{\rotvecne}$ to the corresponding element orientation~$\triadne$:
\begin{equation}
\label{eq:triad_offset}
\triadne
=
\triad\br[1]{\rotvecne}
\triadOffsete
.
\end{equation}
In the present work, the nodal rotation vectors are chosen to vanish in the reference configuration, i.e.,~$\rotvecnO=\mathbf{0}$.
Evaluating~\eqref{eq:triad_offset} in the reference configuration yields
\begin{equation}
\triadOne
=
\triad\br[1]{\rotvecneO}
\triadOffsete
\quad
\xrightarrow{\rotvecneO=\tnsO}
\quad
\triadOne
=
\triadOffsete
.
\end{equation}

For the interpolation of the orientation field, the element orientations are expressed in terms of the local rotation vectors~$\rotvecneRel$ with respect to the intermediate triad~$\triadRefe$:
\begin{equation}
\rotvecneRel = \rv{\br{{\triadRefe}\tr \triadne}}
.
\end{equation}
The local rotation vectors are interpolated within each element using the same shape functions as for the vector space quantities:
\begin{equation}
\rotvechRel\xisolidDependency = \sumbeamne{\Nbeamn\xisolidDependency \rotvecneRel}.
\end{equation}
The interpolated triad is then obtained by evaluating the local triad corresponding to the interpolated local rotation vector and transforming it to the global frame using the intermediate triad:
\begin{equation}
\label{eq:triad_interpolation}
\triad^\ebeam\xisolidDependency \approx \triadhe\xisolidDependency = \triadRefe \triad\br{\rotvechRel\xisolidDependency}.
\end{equation}

For the discretized weak form, finite element approximations of the material curvature~$\beamCurvatureMathe$ and material angular velocity~$\angularVelocityMathe$ are also required.
These approximations are obtained by inserting the interpolated triad field~\eqref{eq:triad_interpolation} into the definitions of the material curvature~\eqref{eq:beamCurvatureMat} and material angular velocity~\eqref{eq:angular_velocity_mat}, together with the definition of the multiplicative infinitesimal rotation~\eqref{eq:mult_variation}.
For further details of the derivation, the reader is referred to~\cite{Jelenic1999}.
The resulting expressions are given by
\begin{align}
\beamCurvatureMat^\ebeam\xisolidDependency &\approx
\beamCurvatureMathe\xisolidDependency
=
\Ttrans\trinv \br{\rotvechRel \xisolidDependency}
\dds{{\rotvechRel}}
\xisolidDependency ,
\\
\angularVelocityMat^\ebeam\xisolidDependency &\approx
\angularVelocityMathe\xisolidDependency
=
\Ttrans\trinv \br{\rotvechRel \xisolidDependency}
\rotvechReldt
\xisolidDependency
.
\end{align}
The same interpolation procedure is employed for the triad field in the reference configuration~$\triadOhe$ and the corresponding reference curvature~$\beamCurvatureOMathe$.

Finally, the interpolation of the spin vector~$\drotmulthe$ has to be defined.
For a Bubnov--Galerkin discretization, the interpolation of the spin vector follows from the variation of~\eqref{eq:triad_interpolation}, the definition of the multiplicative infinitesimal rotation~\eqref{eq:mult_variation}, and subsequent algebraic manipulations, cf.~\cite{Jelenic1999}.
The resulting interpolation functions depend nonlinearly on the primary rotation variables.
Consequently, they would have to be re-evaluated for every new configuration and their dependence on the rotation variables would have to be taken into account for a consistent linearization.
To avoid this additional computational complexity, we adopt the Petrov--Galerkin approach proposed in~\cite{Jelenic1999}, in which the nodal spin vectors~$\drotmultne$ are interpolated using standard Lagrangian finite element shape functions:
\begin{align}
\drotmult^\ebeam \xisolidDependency
\approx
\drotmulthe\xisolidDependency
=
\sumbeamne{
\Nbeamn\xisolidDependency
\drotmultne
}
.
\end{align}

\subsubsection{Construction of the intermediate triad}
\label{sec:discretization_intermediate_triad}

As discussed in the previous subsection, the interpolation of the orientation field requires the introduction of an intermediate triad~$\triadRefe$ for each finite element.
Although the intermediate triad is not unique, it must be defined in an objective manner in order to preserve the objectivity of the resulting interpolation.

The simplest choice is to use the nodal triad of one element node as the intermediate triad.
However, this approach is only applicable if the finite element contains a node that is invariant with respect to the local node numbering, such as the center node of a \quadFE{9} element.
Otherwise, the resulting interpolation remains objective but depends on the chosen local node numbering, which is undesirable.
For elements without a numbering-invariant node, e.g., \quadFE{4} elements, we construct the intermediate triad from the normalized average of the nodal unit quaternions associated with the element nodes.
The normalized average of unit quaternions is objective and independent of the local node numbering, and so is the resulting intermediate triad.

\subsection{Stabilization}
\label{sec:discretization_stabilization}

To improve the numerical stability of the space-time formulation, we employ the residual-based time-upwind stabilization proposed in~\cite{Langer2016} and later applied to linear elasticity in~\cite{Xu2025} and nonlinear elasticity in~\cite{Schuss2023}.
This stabilization can be interpreted as a space-time analogue of the streamline-upwind/Petrov--Galerkin (SUPG) method~\cite{Brooks1982}, with the temporal direction playing the role of the streamline direction.
The finite element formulation may exhibit spurious numerical oscillations which can be interpreted as nonphysical information propagating backward in time.

To suppress these nonphysical modes, the stabilization is introduced by enriching the discrete test functions according to
\begin{align}
\delta\placeholder_h \rightarrow \delta\placeholder_h + \theta h \pfrac{\br{\delta\placeholder_h}}{t}
,
\end{align}
where~$h$ denotes the characteristic element size in the temporal direction and~$\stabil>0$ is a dimensionless stabilization parameter.
The additional term introduces an upwind contribution in the temporal direction, resulting in a Petrov--Galerkin formulation.
It should be noted that the interpolation of the spin vector~$\drotmulth$ already follows a Petrov--Galerkin approach, as discussed in \Cref{sec:discretization_orientations}.
The time-upwind stabilization presented here constitutes an additional Petrov--Galerkin modification and is applied to all test functions of the mixed formulation, i.e.,~$\drbeamh$, $\drotmulth$, $\dvBeamh$, and~$\dangularVelocityMath$.

\subsection{Global system of equations}
\label{sec:discretization_global_system}

The global system of equations is obtained by inserting the stabilized finite element approximations presented in \Cref{sec:discretization_vectorspace,sec:discretization_orientations,sec:discretization_stabilization} into the weak form~\eqref{eq:space_time_weak_form}, yielding the discretized weak form~$\dActMixed \approx \dActMixed_h$.
Since the discretized weak form is linear in the independent variations, these can be factored out, resulting in the global residual vector~$\Rbeam \in \R{12 \nbeamn}$:
\begin{equation}
\label{eq:space_time_weak_form_discretized}
\dActMixed_h =
\vector{\dQbeampos\tr & \dQbeamrot\tr & \dQbeamVel\tr & \dQbeamAngularVel\tr}
\underbrace{\vector{
    \RbeamPos \br{\Qbeampos, \Qbeamrot, \QbeamVel} \\
    \RbeamRot \br{\Qbeampos, \Qbeamrot, \QbeamAngularVel} \\
    \RbeamVel \br{\Qbeampos, \QbeamVel} \\
    \RbeamAngularVel \br{\Qbeamrot, \QbeamAngularVel}
}}_{
    \Rbeam\br{\Qbeampos, \Qbeamrot, \QbeamVel, \QbeamAngularVel}
}
= 0
.
\end{equation}
Here,~$\Qbeampos, \Qbeamrot, \QbeamVel, \QbeamAngularVel \in \R{3\nbeamn}$ denote the global vectors of the primary variables.
The global vectors are obtained by assembling the nodal unknowns, e.g., $\Qbeampos = \brr[0]{ \rbeamni{}_1\tr, \dots{}, \rbeamni{}_{\nbeamn}\tr }\tr$.
The vectors containing the variations are denoted by~$\dQbeampos, \dQbeamrot, \dQbeamVel, \dQbeamAngularVel \in \R{3\nbeamn}$ and are defined analogously.
Finally,~$\RbeamPos, \RbeamRot, \RbeamVel, \RbeamAngularVel \in \R{3\nbeamn}$ denote the global residual vectors.
Since the variations are arbitrary, the global nonlinear system of equations follows as
\begin{equation}
    \label{eq:discretized_system}
\Rbeam\br{\Qbeampos, \Qbeamrot, \QbeamVel, \QbeamAngularVel} = \vvO
.
\end{equation}

To solve~\eqref{eq:discretized_system}, Newton-type solution algorithms require the consistent linearization of the global residual vector~\eqref{eq:space_time_weak_form_discretized}, yielding the global tangent matrix~$\Kbeam \in \R{12 \nbeamn \times 12 \nbeamn}$:
\begin{equation}
\lin{
\vector{
    \RbeamPos \\
    \RbeamRot \\
    \RbeamVel \\
    \RbeamAngularVel
}
}
=
\underbrace{
    \matrix{
    \QbeamPosPos & \QbeamPosRot & \QbeamPosVel & \matO \\
    \QbeamRotPos & \QbeamRotRot & \matO & \QbeamRotAngularVel \\
    \QbeamVelPos & \matO & \QbeamVelVel & \matO \\
    \matO & \QbeamAngularVelRot & \matO & \QbeamAngularVelAngularVel
    }
}_{
    \Kbeam
}
\vector{
   \DQbeampos \\ \DQbeamrot \\ \DQbeamVel \\ \DQbeamAngularVel
}
.
\end{equation}
Here,~$\mat{K}_{\placeholder\placeholderStar} \in \R{3\nbeamn\times 3\nbeamn}$ denotes the Jacobian of the residual vector~$\RbeamPlace$ with respect to the primary variable~$\QbeamPlaceStar$, i.e.,~$\mat{K}_{\placeholder\placeholderStar} = \pfracinline{\vv{r}_{\placeholder}}{\vv{d}_{\placeholderStar}}$.
In practice, all Jacobians are evaluated using the automatic differentiation software package AceGen, cf.~\cite{Korelc2016}.
The increments of the primary variables are denoted by~$\DQbeampos, \DQbeamrot, \DQbeamVel, \DQbeamAngularVel \in \R{3\nbeamn}$.
The updates from iteration~$j$ to iteration~$j+1$ of the vector space variables~$\Qbeampos$, $\QbeamVel$ and~$\QbeamAngularVel$ are performed in an additive manner, i.e.,~$\QbeamPlace^{j+1} = \QbeamPlace^{j} + \DQbeamPlace$.
For the rotational degrees of freedom, two update strategies are available.
A common approach is to solve for multiplicative nodal rotation increments~$\DQbeamrotMult = \brr[0]{ \Drotmultn{}_1\tr, \dots{}, \Drotmultn{}_{\nbeamn}\tr }\tr$ and update the nodal rotation vectors according to~$\rotvecn^{j+1} = \nl(\triad(\Drotmultn_i) \triad(\rotvecn^{j}))$, cf.~\cite{Jelenic1999,Meier2019}.
In the present work, however, we solve for the additive rotation vector increments~$\DQbeamrot = \brr[0]{ \Drotvecni{}_1\tr, \dots{}, \Drotvecni{}_{\nbeamn}\tr }\tr$, such that the rotation degrees of freedom are updated additively, i.e.,~$\Qbeamrot^{j+1} = \Qbeamrot^{j} + \DQbeamrot$.
To maintain a principal representation of the nodal rotations, each nodal rotation vector is mapped after every update such that its norm, i.e., the rotation angle~$\rotvecnorm$, lies within the interval~$[0,\pi]$.

\subsection{Space-time mesh generation}
\label{sec:discretization_mesh_generation}

\begin{figure}
  \centering
  \includegraphics[scale=1]{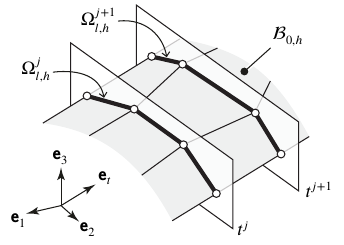}
  \caption{Illustration of the generation of a tensor-product space-time mesh.}
  \label{fig:discretization_mesh_generation}
\end{figure}

An important topic in the context of space-time finite element methods is the generation of 4D space-time meshes.
In particular, we aim to allow for spatial beam configurations whose material domain may vary over time, as required, e.g., for the sliding beam problem considered in \Cref{sec:examples:spaghetti}.
In this work, we assume a tensor-product topology constructed from spatial beam meshes at discrete time levels, i.e., the space-time surface mesh is obtained by connecting compatible beam (curve) meshes at discrete time levels~$t^j$.
The procedure is illustrated in \Cref{fig:discretization_mesh_generation} for a discretization using four-noded~\quadFE{4} elements.
In a first step, the reference space-time surface~$\domainSpaceTimeO$ is sliced at the discrete time levels, resulting in a sequence of spatial beam configurations~$\domainArcLengthi$.
Each spatial configuration is subsequently discretized into beam finite elements, yielding the spatial discretizations~$\domainArcLengthih$.
The nodes of these spatial discretizations at different time levels are not required to correspond to the same material points.
This allows the spatial discretization to vary over time and enables the representation of time-dependent material domains.
However, constructing a tensor-product space-time mesh requires the topology and the number of nodes of the spatial discretization to remain identical at every time level.
Finally, corresponding spatial elements at consecutive time levels are connected in the temporal direction to obtain the space-time finite element mesh~$\domainSpaceTimeOh$.
By construction, one of the two parameter directions of the resulting space-time surface elements follows the spatial beam elements and is therefore aligned with the beam arc-length direction.
The second parameter direction connects corresponding points at consecutive time levels and, in general, need not coincide with the physical time direction.

The space-time mesh generation described above is implemented in the open-source beam finite element pre-processor BeamMe~\cite{BeamMe}.
Its space-time module enables the straightforward generation of 4D space-time meshes for generally curved~3D beam structures.
For each node~$\indexbeam$, the mesh stores the reference position~$\rbeamOn$, the rotation vector~$\rotvecbeamOn$ describing the reference cross section orientation, and the corresponding time coordinate~$\tbeamn$.

\subsection{Numerical integration}
\label{sec:discretization_integration}

Numerical integration over the space-time finite elements is performed using tensor-product Gaussian quadrature.
The integration order can therefore be chosen independently in each parameter direction.
As discussed in \Cref{sec:discretization_mesh_generation}, one parameter coordinate is always aligned with the beam arc-length direction, allowing the quadrature rules in the beam arc-length and second parameter directions to be chosen independently.
To alleviate locking effects, contributions of internal forces to the residual vector and tangent stiffness matrix are evaluated using reduced Gaussian quadrature in the beam arc-length direction, as commonly employed in geometrically exact beam formulations, cf.~\cite{Simo1985,Jelenic1999,Meier2019}.
Full Gaussian quadrature is employed in the second parameter direction.
In the present work, full integration employs $p+1$ integration points, where $p$ is the polynomial degree of the interpolation functions, whereas reduced integration uses $p$ integration points, i.e., one point less.
Full Gaussian quadrature in both parameter directions is applied to the inertia terms and contributions of external loads.

\section{Numerical examples}
\label{sec:numerical_examples}

In this section, we present several numerical examples to demonstrate the capabilities of the proposed space-time formulation.
In \Cref{sec:examples:curved_beam}, we consider a general pre-curved beam and show that the proposed formulation converges to the same solution as classical time-stepping schemes.
Furthermore, we demonstrate objectivity and highlight the effect of space-time stabilization.
In \Cref{sec:examples:free_floating_l}, an L-shaped beam is investigated to illustrate the applicability of the formulation to rigid joints between beams.
In \Cref{sec:examples:spaghetti}, we demonstrate that the proposed space-time formulation can be directly applied to problems with varying material domains by considering the sliding spaghetti problem.

The space-time finite element formulation results in a highly nonlinear system of equations, which may lead to convergence difficulties of the \nr method.
Alternative solution strategies, such as the modified Newton method, have been successfully applied to space-time problems, cf.~\cite{Schuss2023}.
In the present work, we employ a load-stepping approach to improve the robustness of the nonlinear solution procedure.
To this end, the external loads acting on the structure are scaled by a load factor~$\lambda \in [0,1]$, where $\lambda=0$ corresponds to the unloaded configuration and $\lambda=1$ to the prescribed final load.
For each value of~$\lambda$, the resulting nonlinear system of equations is solved using the \nr method, with the converged solution from the previous load step serving as the initial guess for the current one.

The numerical examples are set up with BeamMe~\cite{BeamMe} and simulated using AceGen and AceFEM~\cite{Korelc2016}.
Reference solutions obtained with classical time-stepping schemes are computed using the open-source multiphysics research code 4C~\cite{FourCWebsite}.
The \sr beam elements in 4C are based on the formulation presented in~\cite{Jelenic1999}.
Both first- and second-order spatial discretizations are considered, referred to as \cjShortLine{2} and \cjShortLine{3}, respectively.
For the temporal discretization, the Lie-group generalized-$\alpha$ time integration scheme is employed, cf.~\cite{Bruels2010,Bruels2012}.
The parameters are chosen as $\alpha_m = 0.5$, $\alpha_f = 0.5$, $\beta = 0.25$, and $\gamma = 0.5$, corresponding to a nondissipative scheme with spectral radius $\rho_\infty = 1$.

\subsection{Curved beam}
\label{sec:examples:curved_beam}

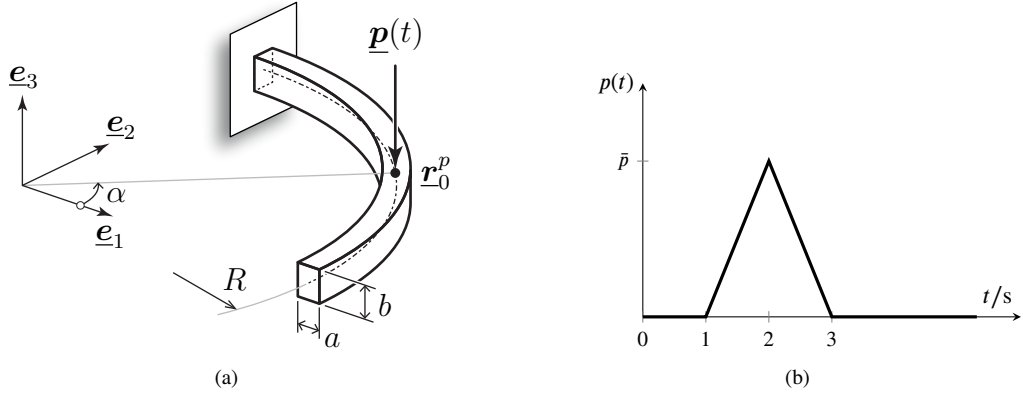
\begin{figure}
	\centering
	\subfigure[]{\label{fig:examples_convergence_problem_problem}\includegraphics[scale=1]{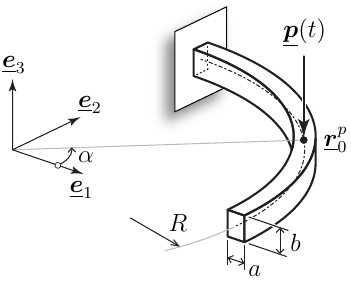}}
	\hfil
	\subfigure[]{\label{fig:examples_convergence_problem_load}
	    \tikzsetnextfilename{examples_convergence_problem_load}
		\begin{tikzpicture}
		\begin{axis}[
			clip=false,
			width=5.cm, 
			height=3.090169943749474cm, 
			scale only axis, 
			xlabel=$t/\unit{s}$, 
			ylabel=$p(t)$,
			axis x line=middle,
			axis y line=middle,
			ylabel style={
				at={(current axis.above origin)},
				anchor=east
			},
			ytick=\empty, 
			xtick=\empty, 
			extra x ticks={0,1,2,3}, 
			extra y ticks={1},
			extra y tick labels={$\bar{p}$}, 
			xmax=6,
			ymax=1.5
		]

		\addplot+[mark=none, very thick, black] coordinates {
			(0.0, 0.0)
			(1.0, 0.0)
			(2.0, 1.0)
			(3.0, 0.0)
			(5.3, 0.0)
		};
		\end{axis}
		\end{tikzpicture}
	}
	\caption{Problem setup of the curved beam example. \subref{fig:examples_convergence_problem_problem} Beam geometry, cross-section shape and boundary conditions. \subref{fig:examples_convergence_problem_load} Time-dependent load applied at the midpoint of the beam centerline.}
	\label{fig:examples_convergence_problem}
\end{figure}

In this first example, we consider an initially curved beam whose stress-free centerline configuration follows a quarter circle, \cf\Cref{fig:examples_convergence_problem_problem}.
The radius~$R$ of the quarter circle is~$\unit[1]{m}$.
The beam has a non-symmetric rectangular cross section with dimensions~$a=\unit[0.075]{m}$ and~$b=\unit[0.1]{m}$, Young's modulus~$\Ebeam=\unit[100]{N/m^2}$, Poisson's ratio~$\nubeam = 0$, and density~$\density=\unit[1]{kg/m^3}$.
One end of the beam is fully clamped, while the other end is free.
A load~$\tns{p}\br{t} = \vector{0, 0, -p\br{t}}\tr$ acting in the negative~$\e3$-direction is applied at the midpoint~$\rbeamO^p$ of the beam centerline, \ie,~$\rbeamO^p = R \vector{\cos{\alpha}, \sin{\alpha}, 0}\tr$ with~$\alpha = \pi/4$.
The magnitude of the load increases linearly from~$p=\unit[0]{N}$ at $t=\unit[1]{s}$ to~$p=\bar{p}=\unit[0.002]{N}$ at~$t=\unit[2]{s}$, and then decreases linearly back to~$p=\unit[0]{N}$ at~$t=\unit[3]{s}$, \cf\Cref{fig:examples_convergence_problem_load}.
We investigate this problem over the time interval $t \in [0, \unit[10]{s}]$.
Until $t=\unit[1]{s}$ no load is applied to the system and the analytical solution is the beam at rest.
We include this rest phase in the numerical simulations to better assess the impact of the stabilization, following a similar approach to that in~\cite{Schuss2023}.
The load on the beam is applied in~$4$ load steps.

\begin{figure}
	\centering
	\subfigure[$t=1.25\,\unit{s}$]{\includegraphics[scale=0.24]{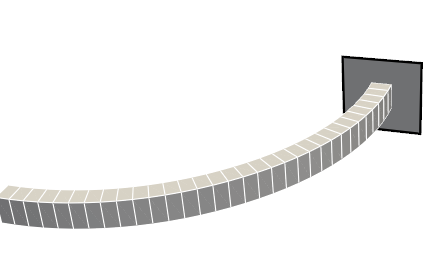}}
    \hfill
	\subfigure[$t=2.5\,\unit{s}$]{\includegraphics[scale=0.24]{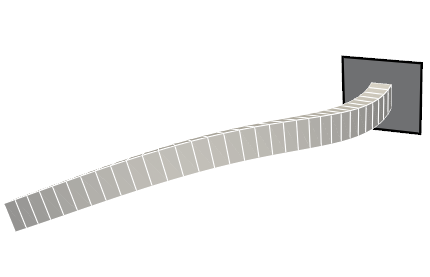}}
    \hfill
	\subfigure[$t=3.75\,\unit{s}$]{\includegraphics[scale=0.24]{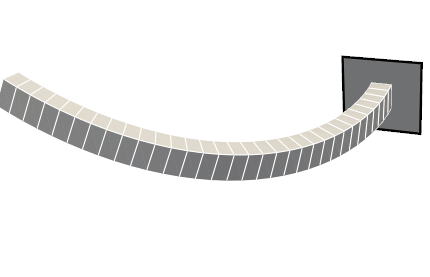}}
    \hfill
	\subfigure[$t=5\,\unit{s}$]{\includegraphics[scale=0.24]{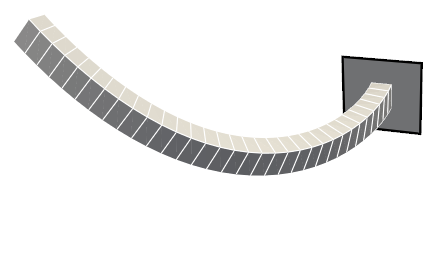}}
    \\
	\subfigure[$t=6.25\,\unit{s}$]{\includegraphics[scale=0.24]{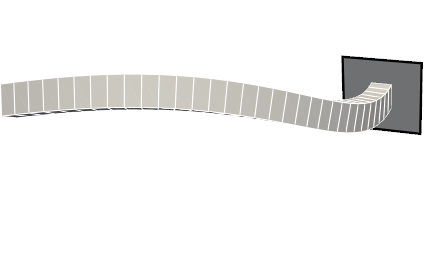}}
    \hfill
	\subfigure[$t=7.5\,\unit{s}$]{\includegraphics[scale=0.24]{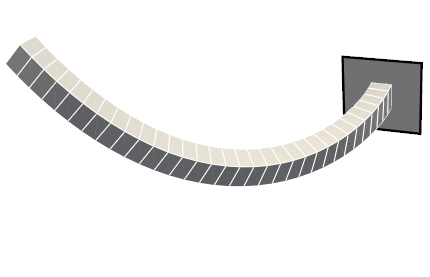}}
    \hfill
	\subfigure[$t=8.75\,\unit{s}$]{\includegraphics[scale=0.24]{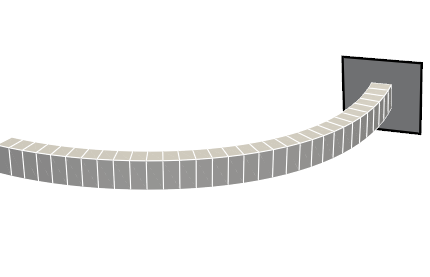}}
    \hfill
    \subfigure[$t=10\,\unit{s}$]{\includegraphics[scale=0.24]{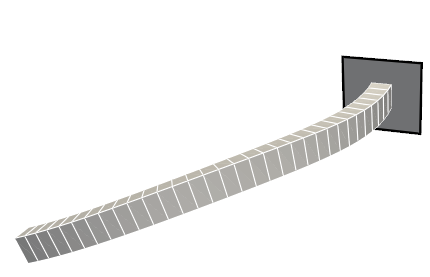}}
	\caption{Snapshots of the deformed space-time beam at different time instances.
    For visualization purposes, the beam is shown as a~3D object with its actual cross section.
    }
	\label{fig:examples_convergence_snapshots}
\end{figure}

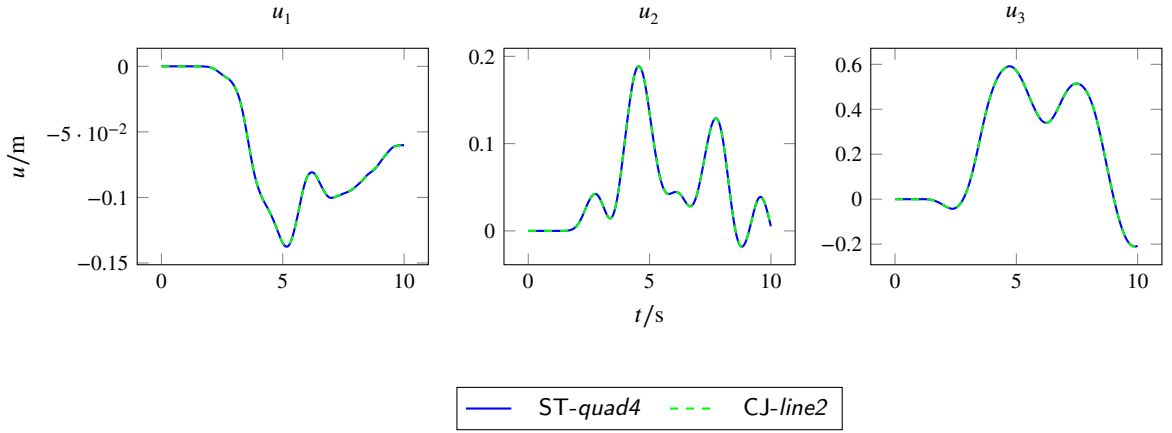
\begin{figure}
	\centering
	\input{figures/examples_convergence_qualitative_displacement.tex}
	\caption{Tip displacements of the curved beam example computed using the proposed space-time formulation (ST-\quadFE{4}) and a standard time-stepping formulation (\cjShortLine{2}).}
	\label{fig:examples_convergence_displacement}
\end{figure}

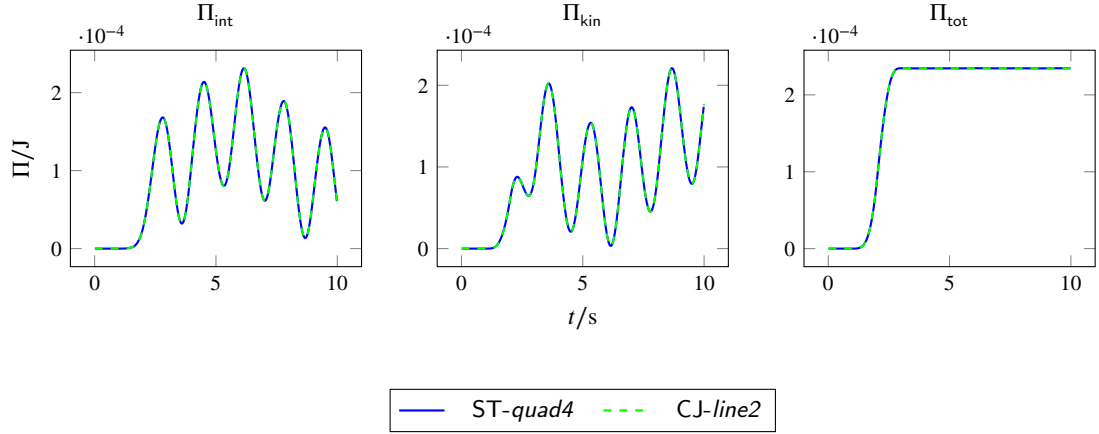
\begin{figure}
	\centering
	\input{figures/examples_convergence_qualitative_energy.tex}
	\caption{System energies of the curved beam example computed using the proposed space-time formulation (ST-\quadFE{4}) and a standard time-stepping formulation (\cjShortLine{2}).}
	\label{fig:examples_convergence_energy}
\end{figure}

Snapshots of the deformed configuration of the curved beam are presented in \Cref{fig:examples_convergence_snapshots}.
The results are obtained using the proposed space-time formulation with bilinear \quadFE{4} elements,~$n_s = 32$ elements in the spatial direction and~$n_t = 320$ elements in the temporal direction.
The stabilization parameter is chosen as~$\stabil = 0.1$. For a discussion of this choice, we refer to \Cref{sec:examples:curved_beam:uniform,sec:examples:curved_beam:stabilization}.
The tip displacement in all three spatial directions is compared in \Cref{fig:examples_convergence_displacement} for the proposed space-time formulation and the classical time-stepping \cjShortLine{2} formulation.
The classical time-stepping formulation employs the same spatial discretization and temporal resolution as the proposed space-time formulation.
In particular, the number of time steps in the classical formulation is chosen equal to the number of temporal elements used in the space-time discretization.
The results obtained with the two formulations are in excellent agreement.
In \Cref{fig:examples_convergence_energy}, we also plot and compare the kinetic, elastic, and total system energies over time.
Once again, the results from the two numerical formulations closely match each other.

In \Cref{sec:examples:curved_beam:uniform,sec:examples:curved_beam:stabilization,sec:examples:curved_beam:objectivity}, this example is used to further investigate the proposed space-time formulation, with particular focus on temporal and spatial convergence behavior, the effect of space-time stabilization, as well as the objectivity of the formulation.

\subsubsection{Convergence for uniform refinement}
\label{sec:examples:curved_beam:uniform}

To investigate the numerical error and the convergence behavior of the proposed space-time formulation, we perform a uniform refinement study in both space and time.
To this end, we consider the curved beam example with different discretization levels, where the number of elements in the spatial direction~$n_s$ and the number of elements in the temporal direction~$n_t$ are increased simultaneously with the same refinement factor~$r$.
The baseline discretization consists of~$1$ element in the spatial direction and~$10$ elements in the temporal direction.
Thus, the discretization corresponding to refinement factor~$r$ is given by~$n_s = r$ and~$n_t = 10 r$.
The numerical error is quantified by the relative~$L^2$ displacement error over the complete space-time surface:
\begin{equation}
\label{eq:examples:curved_beam:error}
\relErrorDisp
=
\sqrt{
    \frac{
        \intSpaceTimeFixedMaterial{
            (\ubeam[\text{h}] - \ubeam[\text{ref}])\tr(\ubeam[\text{h}] - \ubeam[\text{ref}])
        }
    }
    {
        \intSpaceTimeFixedMaterial{
            \ubeam[\text{ref}]\tr\ubeam[\text{ref}]
        }
    }
}
.
\end{equation}
Here,~$\ubeam[\text{h}]$ denotes the displacement field of the discretization under consideration, whereas~$\ubeam[\text{ref}]$ denotes the reference solution computed with the classical \cjShortLine{3} formulation using a highly refined discretization consisting of~$n_s = 256$ spatial elements and~$n_t = 81920$ time steps.
In practice, the integral in \Cref{eq:examples:curved_beam:error} is evaluated by Gaussian quadrature over the space-time surface.
For solutions obtained with a classical time-stepping formulation, the corresponding space-time representation is constructed by linear interpolation between consecutive time steps.

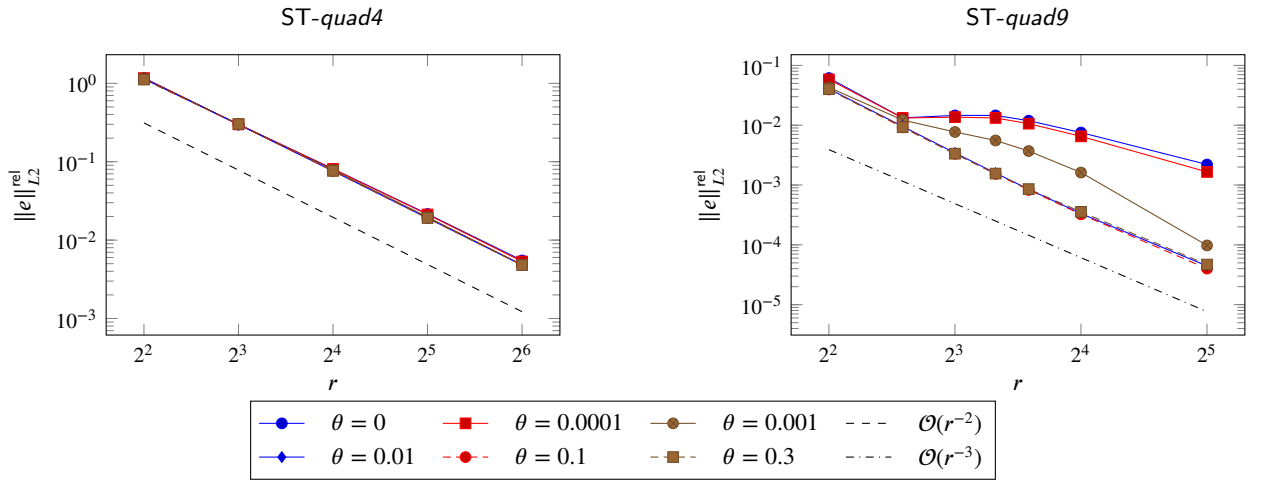
\begin{figure}
	\centering
	\input{figures/examples_convergence_uniform_quad4.tex}
    \hfill
    \input{figures/examples_convergence_uniform_quad9.tex}
    \tikzset{external/export next=false}
    \tikzsetnextfilename{examples_convergence_uniform_legend}
    \pgfplotslegendfromname{examples_convergence_uniform_legend}
	\caption{Relative $L^2$ displacement error under uniform refinement in space and time for the proposed space-time formulation: first-order ST-\quadFE{4} elements~(left) and second-order ST-\quadFE{9} elements~(right). Results are shown for different values of the stabilization parameter~$\stabil$.}
	\label{fig:examples_convergence_uniform}
\end{figure}

The results of the uniform mesh refinement study are presented in \Cref{fig:examples_convergence_uniform}.
The left plot shows the convergence behavior of the proposed space-time formulation using bilinear \quadFE{4} elements, whereas the right plot presents the results for biquadratic \quadFE{9} elements.
The simulations are performed for stabilization parameters~$\stabil \in \{0,0.0001,0.001,0.01,0.1,0.3\}$.
For the ST-\quadFE{4} elements, the expected second-order convergence is observed.
Moreover, the convergence is practically independent of the stabilization parameter.
The ST-\quadFE{9} elements exhibit a more complex convergence behavior.
Here, the stabilization parameter has a significant influence on both the convergence rate and the error magnitude.
For small stabilization parameters,~$\stabil \leq 0.001$, the convergence rate is reduced and the error is noticeably larger than for larger values of~$\stabil$.
In contrast, for~$\stabil \geq 0.01$, the expected third-order convergence is recovered.
Overall, the results demonstrate that the proposed space-time formulation converges to the same solution as the corresponding classical time-stepping formulation.
Furthermore, they highlight the importance of the space-time stabilization, which is investigated in more detail in \Cref{sec:examples:curved_beam:stabilization}.

\begin{figure}
	\centering
	\input{figures/examples_convergence_uniform_all.tex}
	\caption{Comparison of the convergence behavior of first- and second-order space-time elements with first- and second-order finite elements in space with classical time-stepping formulations.}
	\label{fig:examples_convergence_uniform_all}
\end{figure}
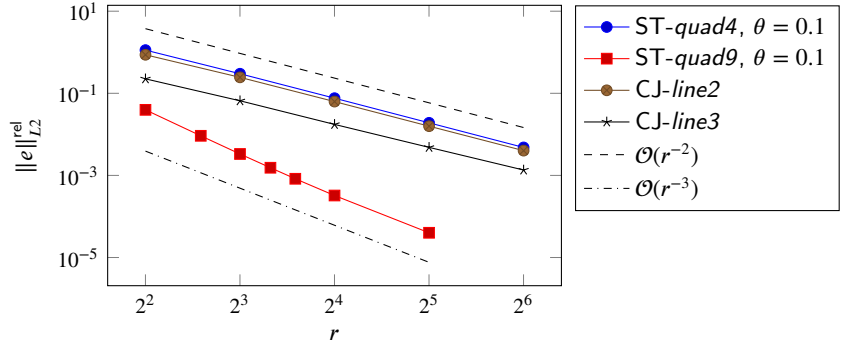

In \Cref{fig:examples_convergence_uniform_all}, the convergence behavior of first- and second-order space-time finite elements (ST-\quadFE{4} and ST-\quadFE{9}) with~$\stabil = 0.1$ is compared with first- and second-order finite elements in space (\cjShortLine{2} and \cjShortLine{3}) combined with classical time integration.
The ST-\quadFE{4} and \cjShortLine{2} formulations exhibit very similar error levels and both achieve the expected second-order convergence.
Although the \cjShortLine{3} formulation employs second-order finite elements in space, the overall convergence order is limited by the generalized-$\alpha$ time integration scheme, resulting in an overall second-order convergence rate.
However, due to the second-order approximation in spatial direction, the error level is lower than for the ST-\quadFE{4} formulation.
In contrast, the ST-\quadFE{9} formulation is second-order in both space and time and therefore achieves the expected third-order convergence and the overall lowest error levels.

\subsubsection{Effect of space-time stabilization}
\label{sec:examples:curved_beam:stabilization}

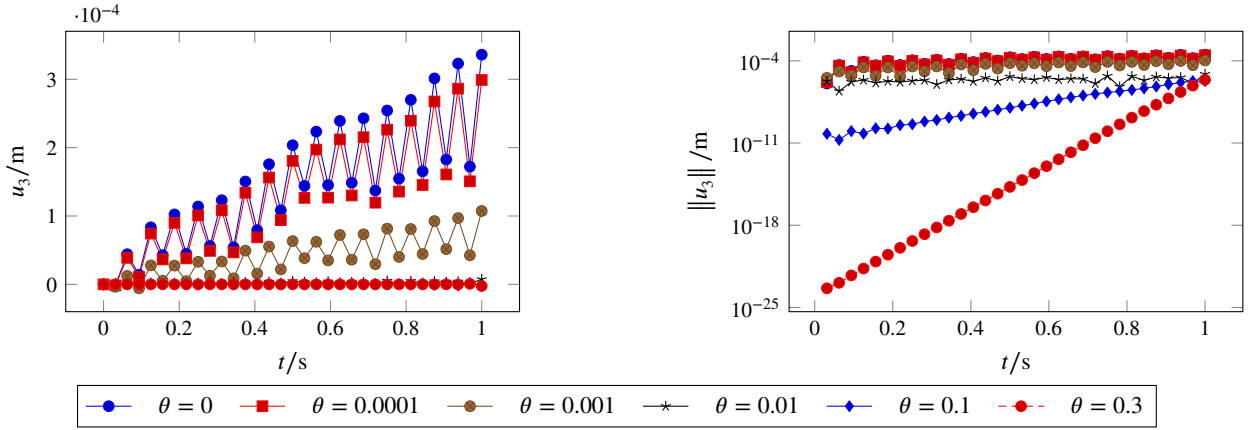
\begin{figure}
	\centering
	\input{figures/examples_convergence_stabilization.tex}
	\caption{Space-time solution for the curved beam example in the interval $t \in [0, \unit[1]{s}]$ for different values of the stabilization parameter $\stabil$.
    Displacement~$u_3$ in the~$\ez$-direction at the point of application of the load~$\rbeamO^p$ (left) and
    absolute value of~$u_3$ on a logarithmic scale (right).
    The markers shown correspond to the nodes in time direction at~$\rbeamO^p$.}
	\label{fig:examples_convergence_stabilization}
\end{figure}

In the previous subsection, it was shown that the space-time stabilization can improve both the accuracy of the solution and the convergence behavior.
The effect of the stabilization is now investigated in more detail, following a similar approach to that proposed in~\cite{Schuss2023}.
As explained in \Cref{sec:examples:curved_beam}, the beam is initially at rest and no external load is applied during the time interval~$t \in [\unit[0]{s}, \unit[1]{s}]$.
Consequently, the exact solution is the trivial solution, i.e., the beam remains at rest throughout this time interval.
To assess whether this behavior is accurately reproduced by the space-time formulation, the results are analyzed over the interval~$t \in [\unit[0]{s}, \unit[1]{s}]$, while the total simulation time is still~$\unit[10]{s}$.
The space-time discretization employs ST-\quadFE{4} elements with~$n_s = 8$ spatial elements and~$n_t = 320$ temporal elements.
\Cref{fig:examples_convergence_stabilization}(left) shows the beam displacement~$u_3$ in the~$\ez$-direction at the point of application of the load~$\rbeamO^p$, for different values of the stabilization parameter~$\stabil$.
For all considered stabilization parameters, the exact trivial solution is not recovered.
One can see nonzero displacements oscillating between nodal values.
This behavior can be interpreted as spurious information propagating backward in time through the space-time mesh, which is unphysical.
It can clearly be seen that the stabilization parameter has a significant influence on this effect, with smaller values of~$\stabil$ leading to larger oscillations of the displacement field.
\Cref{fig:examples_convergence_stabilization}(right) further illustrates this behavior by plotting the absolute value of~$u_3$ on a logarithmic scale.
The results indicate that the space-time stabilization acts as a damping mechanism for these spurious backward-traveling modes.
Based on these findings, together with the convergence results presented in the previous subsection, a stabilization parameter of~$\stabil = 0.1$ is adopted for all subsequent simulations.
It is important to note that the choice of the stabilization parameter is problem-dependent.
A rigorous mathematical analysis of the stabilization is beyond the scope of the present work and will be the subject of future research.
Nevertheless, it is worth noting that the results reported in~\cite{Schuss2023} lead to similar conclusions and likewise recommend a stabilization parameter of~$\stabil = 0.1$.

\subsubsection{Objectivity of the space-time formulation}
\label{sec:examples:curved_beam:objectivity}

As a final aspect of the curved beam analysis, we investigate the objectivity of the proposed space-time finite element formulation.
To this end, we compare solutions obtained from initial configurations that only differ by a rigid body rotation~$\rotMatArbitrary$.
The rigid body rotation is applied to the initial beam position and orientation as well as to the external loads, i.e.,~$\rbeamO^* = \rotMatArbitrary \rbeamO$, $\triadO^* = \rotMatArbitrary \triadO$, and~$\tns{p}^* = \rotMatArbitrary \tns{p}$.
The beam is discretized using \quadFE{4} elements with~$n_s=8$ elements in the spatial direction and~$n_t=320$ elements in the temporal direction, the stabilization parameter is set to~$\stabil=0.1$.

The tip displacements obtained for three different initial configurations are reported in \Cref{tab:examples_objectivity}.
As expected, the absolute tip displacements~$\ubeam$ do not coincide, however, after applying the corresponding back-transformation, i.e.,~$\br[0]{\rotMatArbitrary}\tr\ubeam$, the resulting tip displacements agree up to the nonlinear solver tolerance.
This confirms the objectivity of the proposed space-time formulation.

\begin{table}
    \centering
    \caption{Comparison of tip displacements obtained from initial configurations differing only by a rigid body rotation.}
    \label{tab:examples_objectivity}
    \begin{tabular}{ccc}
        \toprule
        $\rotvecArbitrary$ & $\ubeam / \unit{m}$ & $\br[1]{\triad\tr\br[1]{\rotvecArbitrary}\ubeam} /\unit{m}$  \\
        \midrule
        $\vector{0\\0\\0}$ &
        $\vector{-0.04620057139214\\ -0.01554262752001\\ -0.15866137965134}$ &
        $\vector{-0.04620057139214\\ -0.01554262752001\\ -0.15866137965134}$ \\[3ex]

        $\vector{0\\0\\-\pi/3}$ &
        $\vector{-0.03656059596995\\ \phantom{-}0.03223955473494\\ -0.15866137965134}$ &
        $\vector{-0.04620057139214\\ -0.01554262752001\\ -0.15866137965134}$ \\[3ex]

        $\vector{-\pi/3\\-\pi/3\\-\pi/3}$ &
        $\vector{\phantom{-}0.01172499394235\\ -0.10073581431685\\ -0.13139375818898}$ &
        $\vector{-0.04620057139214\\ -0.01554262752001\\ -0.15866137965134}$ \\
        \bottomrule
    \end{tabular}
\end{table}

\subsection{Free-floating L-shaped beam}
\label{sec:examples:free_floating_l}

\begin{figure}
	\centering
	\includegraphics[scale=1]{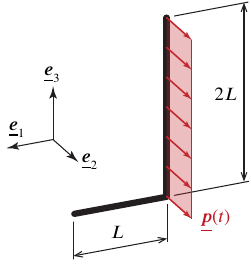}
	\caption{Problem setup of the free-floating L-shaped beam consisting of two beam segments connected by a rigid joint.}
	\label{fig:examples_free_floating_l_problem}
\end{figure}

In this example, we investigate the dynamics of a free-floating L-shaped beam, as shown in \Cref{fig:examples_free_floating_l_problem}.
The beam consists of two straight segments connected by a rigid joint.
The segment in the~$\ex$-direction has a length of~$L$ and the segment in the~$\ez$-direction has a length of~$2L$ with~$L=\unit[1]{m}$.
Both segments have a square cross section with side length~$a=\unit[0.1]{m}$, Young's modulus~$\Ebeam=\unit[100]{N/m^2}$, Poisson's ratio~$\nubeam = 0$, and density~$\density=\unit[1]{kg/m^3}$.
The long segment of the beam is loaded with a time-dependent distributed line load~$\tns{p}\br{t} = \vector{0, p\br{t}, 0}\tr$ acting in the~$\ey$-direction.
The magnitude of the line load increases linearly from~$p\br{t}=\unit[0]{N/m}$ at $t=\unit[0]{s}$ to~$p\br{t}=\unit[0.005]{N/m}$ at~$t=\unit[1]{s}$, and then decreases linearly back to~$p\br{t}=\unit[0]{N/m}$ at~$t=\unit[2]{s}$.
No Dirichlet-type conditions are applied at the boundaries of the spatial domain of the beam.
The dynamic response is computed over the time interval~$t \in [\unit[0]{s}, \unit[10]{s}]$.
This example demonstrates the applicability of the proposed formulation to beam structures composed of multiple beam segments connected by rigid joints, cf.~\Cref{sec:discretization_orientations}.

The problem is discretized using second-order \quadFE{9} space-time finite elements.
The short and long beam segments are discretized with~$n_{s1} = 5$ and~$n_{s2} = 10$ elements in the spatial direction, respectively.
Both beam segments are discretized with~$n_t = 200$ elements in the temporal direction.
A stabilization parameter of~$\stabil = 0.1$ is employed, and the external load is applied in~$10$ load steps.
The results are shown in \Cref{fig:examples_free_floating_l_deformed}.
The left image depicts the deformed space-time surface of the L-shaped beam.
Black curves correspond to constant-time slices through the space-time surface and illustrate the beam configuration at selected time instances.
The corresponding beam configurations are shown on the right.
The displacement history of the joint is compared with the classical time-stepping formulation~\cjShortLine{3} using the same spatial discretization and~1000 time steps in \Cref{fig:examples_free_floating_l_displacement}.
One can see that the results are in very good agreement, showing that the proposed space-time formulation can accurately capture the dynamics of beams with rigid joints.
\begin{figure}
	\centering
    \subfigure[]{%
        \label{fig:examples_free_floating_l_deformed_surface}%
        \includegraphics[scale=0.24]{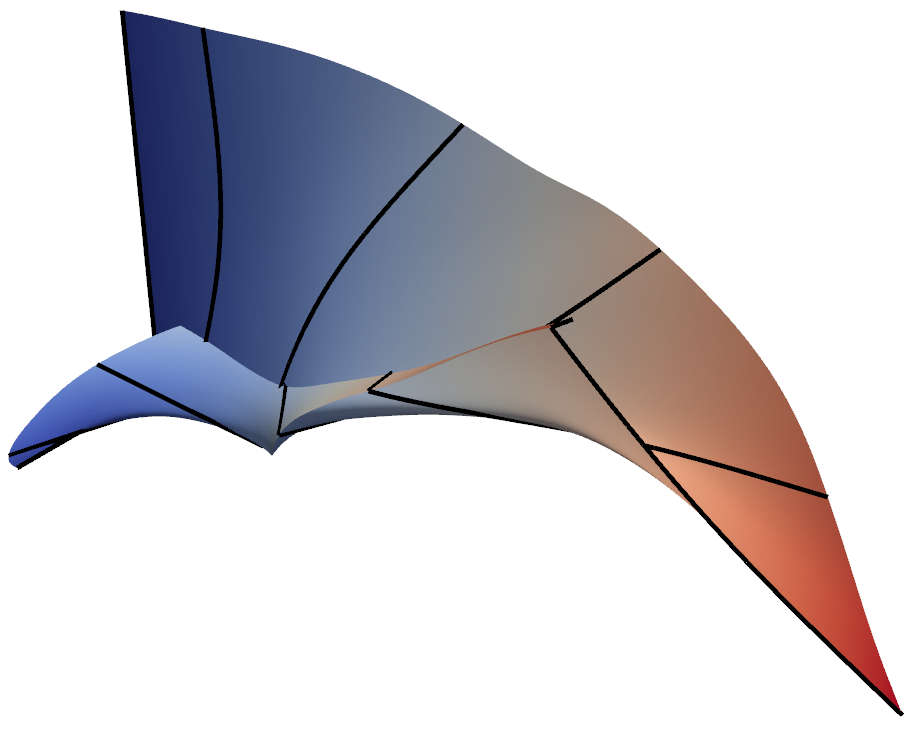}%
    }
    \hfill
    \subfigure[]{%
        \label{fig:examples_free_floating_l_deformed_snapshots}%
        \input{figures/examples_free_floating_l_deformed.tex}%
    }
	\caption{Results of the L-shaped beam. \subref{fig:examples_free_floating_l_deformed_surface} Deformed space-time surface. \subref{fig:examples_free_floating_l_deformed_snapshots} Corresponding beam configurations at selected time instances, for visualization purposes, the beams are shown as~3D objects with their actual cross section.}
	\label{fig:examples_free_floating_l_deformed}
\end{figure}
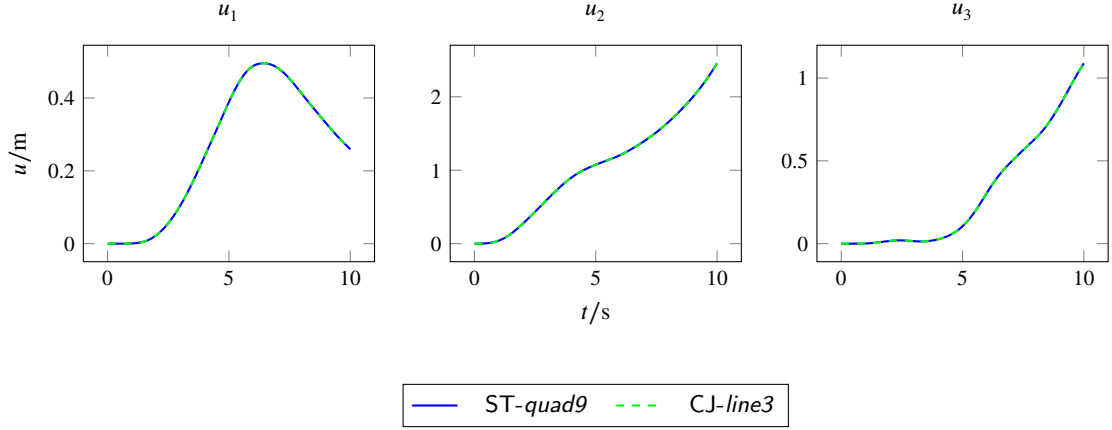
\begin{figure}
	\centering
	\input{figures/examples_free_floating_l_displacement.tex}
    \caption{Displacement history of the joint computed using the proposed space-time formulation (ST-\quadFE{9}) and a standard time-stepping formulation (\cjShortLine{3}).}
	\label{fig:examples_free_floating_l_displacement}
\end{figure}

\subsection{Sliding spaghetti problem}
\label{sec:examples:spaghetti}

\begin{figure}
	\centering
    \subfigure[]{\label{fig:examples_sliding_beam_problem}\includegraphics[scale=1]{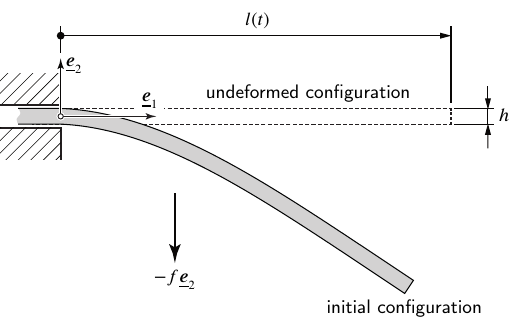}}
    \hfill
    \subfigure[]{\label{fig:examples_sliding_beam_mesh}\input{figures/examples_sliding_beam_mesh.tex}}
	\caption{Sliding spaghetti problem. \subref{fig:examples_sliding_beam_problem} Problem setup with the beam partially inserted into a rigid guide.
    \subref{fig:examples_sliding_beam_mesh} Undeformed space-time finite element mesh in the~$\sbeam$--$t$ plane illustrating the meshing of the time-dependent material domain. For visualization purposes, a coarser mesh than that used in the simulation is shown.}
	\label{fig:examples_sliding_beam}
\end{figure}

In this example, we demonstrate the applicability of the proposed space-time formulation to problems with time-dependent material domains by considering a variant of the sliding spaghetti problem, cf.~\cite{Steinbrecher2017,VuQuoc1995,Humer2013,Humer2020,Carrier1949}.
The problem is based on the sliding spaghetti example presented in~\cite{Steinbrecher2017,Humer2020}.
\Cref{fig:examples_sliding_beam_problem} depicts the problem setup, which consists of a beam that is partially inserted into a rigid guide.
The initially straight beam is subject to a gravity load~$-f \ey$, which results in a downward-bent prestressed initial configuration.
The sliding motion at the guide is fully prescribed, such that the material length of the beam outside of the guide is given by~$l(t)$.
To set the structure smoothly into motion, a piecewise linear acceleration over the time interval~$t \in [0, \timeEnd_a]$ is chosen:
\begin{equation}
    a(t) = 
    \begin{cases} 
        -2 a_0 \frac{t}{\timeEnd_a}, & t \leq \frac{\timeEnd_a}{2} \\
        -2 a_0 \br[1]{1-\frac{t}{\timeEnd_a}}, & \frac{\timeEnd_a}{2} < t \le \timeEnd_a \\
        0, & t > \timeEnd_a
    \end{cases}
.
\end{equation}
Integrating twice, the acceleration translates into the corresponding material length outside the guide:
\begin{equation}
    l(t) = L +
    \begin{cases} 
        -\frac{1}{3} a_0 \frac{t^3}{\timeEnd_a}, & t \leq \frac{\timeEnd_a}{2} \\
        -\frac{1}{4} a_0 \br{2 t - \timeEnd_a} \timeEnd_a, & \frac{\timeEnd_a}{2} < t \le \timeEnd_a \\
        -\frac{a_0}{12 \timeEnd_a} \br{\timeEnd_a^3 - 6 t \timeEnd_a^2 + 12 t^2 \timeEnd_a - 4 t^3}, & t > \timeEnd_a
    \end{cases}
.
\end{equation}
\Cref{tbl:contact:examples:spaghetti_parameter} lists the remaining properties of the problem.
\begin{table}
\centering
\caption{Problem parameters for the sliding spaghetti problem.}
\label{tbl:contact:examples:spaghetti_parameter}
\begin{tabular}{lcr}
\toprule
	initial beam length        &        $L$        &          $\unit[5]{m}$ \\
	beam width                 &        $b$        &          $\unit[1]{m}$ \\
	beam height                &        $h$        &        $\unit[0.2]{m}$ \\
    shear correction factor            & $k$ &                    $5/6$ \\
	Young's modulus            &  $E$  &   $\unit[1000]{N/m^2}$ \\
	Poisson's ratio            & $\nu$ &                    $0$ \\
    density            & $\rho$ &                    $\unit[1000]{N/m^3}$ \\
    gravity load            & $f$ &                    $\unit[200]{N/m}$ \\
    peak acceleration            & $a_0$ &                    $\unit[1]{m/s^2}$ \\
    acceleration time            & $\timeEnd_a$ &                    $\unit[2]{s}$ \\
    total time            & $\timeEnd$ &                    $\unit[5]{s}$ \\
\bottomrule
\end{tabular}
\end{table}

The main challenge of this problem is the treatment of the interaction between the beam and the rigid guide, which acts as a non-material boundary.
Several approaches have been proposed for this purpose, see~\cite{Steinbrecher2017}.
Within a classical Lagrangian finite element formulation, the guide is typically modeled using contact constraints, which considerably increase the computational complexity.
An alternative is the Sliding Beam Formulation~(SBF)~\cite{Humer2013,Humer2020}, a mixed Lagrangian--Eulerian approach in which a non-material coordinate transformation maps the material point located at the opening of the guide to a fixed position in the transformed domain.
This enables the direct application of the boundary conditions at the guide.
However, the derivation of the SBF is rather involved and has so far only been presented for planar beam formulations.
The proposed space-time formulation provides an alternative treatment of the moving material domain.
Since we discretize in both space and time, the considered material domain can vary over the space-time mesh.
This is illustrated in \Cref{fig:examples_sliding_beam_mesh}, where the undeformed space-time mesh is shown in the~$\sbeam$--$t$ plane.
The right boundary~$\boundarySpaceB$ corresponds to the free beam tip.
The material coordinate of the left boundary~$\boundarySpaceA$ varies over time such that the considered material length at time~$t$ corresponds to~$l(t)$.
Consequently, every node on~$\boundarySpaceA$ represents the material point currently located at the guide opening.
The Dirichlet boundary conditions enforcing the sliding motion can therefore be imposed directly on this non-material boundary:
\begin{equation}
    \ubeam = \bar{\ubeam} = \vector{-\br{L-l(t)} \\ 0 \\ 0} \quad \text{on} \quad \boundarySpaceA
    .
\end{equation}
Here, we have to apply a nonzero Dirichlet condition in the~$\ex$-direction, which ensures that each node along~$\boundarySpaceA$ is always located at the opening of the guide.

The problem is discretized using first-order \quadFE{4} space-time finite elements with~$n_s = 20$ and ~$n_t = 2000$ elements in the spatial and temporal direction, respectively.
Since the material domain changes over time while the number of spatial elements remains constant, the element length in the spatial direction varies over time.
Consequently, the spatial discretization becomes finer as the simulated material length of the beam decreases.
This behavior is illustrated in \Cref{fig:examples_sliding_beam_mesh}.
For the sake of clarity, however, a coarser mesh than the one used in the simulation is shown.
A stabilization parameter of~$\stabil = 0.1$ is employed, and the gravity load is applied in~10 load steps.

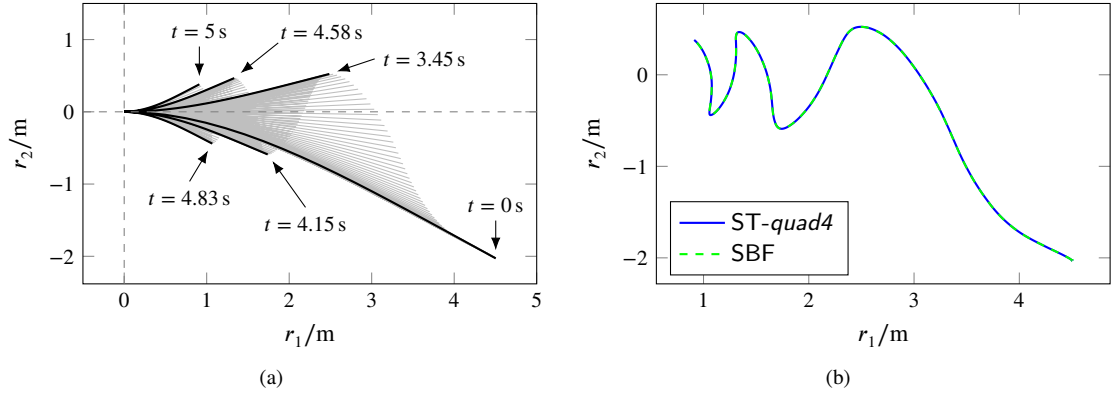
\begin{figure}
	\centering
    \subfigure[]{%
        \label{fig:examples_sliding_beam_results_configuration}%
        \input{figures/examples_sliding_beam_configuration.tex}%
    }
	\subfigure[]{%
		\label{fig:examples_sliding_beam_results_tip_displacement}%
		\input{figures/examples_sliding_beam_tip_displacement.tex}%
	}
	\caption{Results for the sliding spaghetti example. \subref{fig:examples_sliding_beam_results_configuration} Series of deformed beam configurations over the course of motion.
    \subref{fig:examples_sliding_beam_results_tip_displacement} Beam tip trajectory comparing the proposed space-time formulation ST-\quadFE{4} with the Sliding Beam Formulation~(SBF).}
	\label{fig:examples_sliding_beam_results}
\end{figure}

The results are presented in \Cref{fig:examples_sliding_beam_results}.
\Cref{fig:examples_sliding_beam_results_configuration} shows a series of deformed beam configurations over the course of motion.
It can be seen that the oscillation frequency increases over time.
This is expected, since the simulated beam length continuously decreases, resulting in an increase of the natural frequencies.
\Cref{fig:examples_sliding_beam_results_tip_displacement} shows the trajectory of the beam tip and compares it with the SBF.
The SBF model consists of 20 first-order beam finite elements in spatial direction and 2000 time steps are employed for time integration.
The two fundamentally different approaches show excellent agreement, demonstrating the applicability of the proposed space-time formulation to problems with time-dependent material domains.

It is important to note that no modifications to the proposed space-time formulation were required to model this problem.
In the present example, the motion at the guide is prescribed purely kinematically.
In some applications, however, it may also be of interest to consider the dynamics of the beam segment located inside the guide, cf.~\cite{Humer2013,Steinbrecher2017,Humer2020}.
Such problems cannot be treated directly using the present approach.
Nevertheless, the introduction of a suitable coordinate transformation, similar to that employed in the SBF, would enable the proposed space-time formulation to account for the dynamics of the complete beam.

\section{Conclusion}
\label{sec:conclusion}

In this work, we presented a space-time finite element formulation for the transient analysis of geometrically exact shear-deformable beams.
The formulation employs a mixed approach in which the translational and angular velocities are introduced as independent fields, reducing the highest temporal derivatives to first order and allowing velocity-level initial conditions to be imposed directly.
Finite rotations are represented by nodal rotation (pseudo-)vectors and interpolated on the space-time surface using an objective rotation interpolation.
The resulting formulation is applicable to generally curved beam structures and allows for geometric kinks and rigid joints.

The numerical examples demonstrated that the proposed space-time formulation converges to the same solution as classical time-stepping approaches.
Both first- and second-order space-time finite elements showed the expected convergence behavior, while the numerical investigations also highlighted the importance of the time-upwind stabilization for suppressing spurious oscillations associated with information propagating backward in time.
Furthermore, the objectivity of the formulation was demonstrated numerically.
A particular advantage of the space-time setting is the direct representation of time-dependent material domains through the geometry of the space-time surface.
As demonstrated by the sliding spaghetti problem, this allows non-material boundaries to be represented without modifying the underlying beam formulation.

\section*{Acknowledgements}

Sketches in this work have been created using the Adobe Illustrator plug-in LaTeX2AI (\url{https://github.com/isteinbrecher/latex2ai}).

\section*{Declaration of generative AI and AI-assisted technologies in the writing process}

During the preparation of this work, the authors used ChatGPT-5.6 Sol to fix writing errors and to improve the readability and language in some parts of the manuscript.
After using these tools, the authors reviewed and edited the content and take full responsibility for the content of the published article.


\printcredits

\bibliographystyle{cas-model2-names}

\bibliography{literature_filtered,steinbrecher_filtered}

\end{document}

%% file: tex/packages.tex
\usepackage[utf8]{inputenc}

\usepackage[ngerman,american]{babel}

\usepackage{microtype}

\usepackage{amsmath,amsfonts,amssymb}
\usepackage{stix}
\usepackage[scaled=.85]{inconsolata}

\usepackage{graphicx}

\usepackage{xspace}

\usepackage{units}

\usepackage{color}

\usepackage{amsthm}
\newtheorem{remark}{Remark}

\usepackage{booktabs}
\usepackage{longtable}
\usepackage{tabularx}
\usepackage{array}

\usepackage{subfigure}

\usepackage{hyperref}

\usepackage{pdflscape}
\usepackage[figuresright]{rotating}

\usepackage{cleveref}
\crefrangeformat{equation}{(#3#1#4) to~(#5#2#6)}
\crefmultiformat{equation}{(#2#1#3)}{ and~(#2#1#3)}{, (#2#1#3)}{ and~(#2#1#3)}

\usepackage[numbers,sort&compress]{natbib}

\usepackage{xfp}

\usepackage{enumerate} 

\usepackage{tikz}
\usepackage{pgfplots}
\pgfplotsset{compat=1.16}
\usetikzlibrary{arrows.meta}
\usetikzlibrary{external}
\usepgfplotslibrary{groupplots}
\pgfplotsset{
    every axis/.append style={
        tick label style={font=\footnotesize},
        width=6cm,
        height=3.708203928cm
    }
}

\makeatletter
\renewcommand\@pnumwidth{2em}%
\makeatother

\usepackage{ifdraft}

%% file: tex/shortcuts.tex
\usepackage{etoolbox}
\usepackage{xifthen}
\usepackage{bm}
\usepackage{amssymb}
\usepackage{mathtools}
\usepackage[dvipsnames]{xcolor}

\newcommand{\ie}{i.e.\xspace}

\newcommand{\cf}{cf.~}

\newcommand{\sr}{Simo--Reissner\xspace}

\newcommand{\cs}[1][]{cross-section#1\xspace}

\newcommand{\nr}{Newton--Raphson\xspace}
\newcommand{\cosserat}{Cosserat\xspace}

\newcommand{\cjShort}{CJ\xspace}
\newcommand{\cjShortLine}[1]{\cjShort-\emph{line#1}}

\newcommand{\Gexact}{Geometrically exact\xspace}

\newcommand{\quadFE}[1]{\emph{quad#1}\xspace}

\newcommand{\R}[1]{\mathbb{R}^{#1}}

\newcommand{\SO}{SO^3}
\newcommand{\SOtan}{so^3}
\newcommand{\order}[1]{\mathcal O(#1)}

\newcommand{\e}[1]{\tns{e}_{#1}}
\newcommand{\ei}{\tns{e}_{i}}
\newcommand{\ex}{\e{1}}
\newcommand{\ey}{\e{2}}
\newcommand{\ez}{\e{3}}

\newcommand{\eST}[1]{\tnsST{e}_{#1}}
\newcommand{\eSTx}{\eST{1}}
\newcommand{\eSTy}{\eST{2}}
\newcommand{\eSTz}{\eST{3}}
\newcommand{\eSTt}{\eST{t}}

\newcommand{\tnsBr}[1]{\brrr{#1}}

\newcommand{\pfrac}[2]{\frac{\partial #1}{\partial #2}}
\newcommand{\pfracinline}[2]{\partial #1/\partial #2}
\newcommand{\ptfrac}[2]{\frac{\mathrm d #1}{\mathrm d #2}}
\DeclareMathOperator{\lin}{Lin}
\newcommand{\dds}[1]{#1'}
\newcommand{\ddt}[1]{\dot{#1}}
\newcommand{\dddt}[1]{\ddot{#1}}

\newcommand{\tn}[2]{%
\ifnumcomp{#1}{=}{1}{\underline{\boldsymbol{#2}}}
{\ifnumcomp{#1}{=}{2}{\underline{\boldsymbol{#2}}}
{\ifnumcomp{#1}{>}{2}{Higher order tensor not yet implemented!}
{Wrong tensor order given}%
}}}
\newcommand{\tns}[1]{\tn{1}{#1}}
\newcommand{\tnss}[1]{\tn{2}{#1}}
\newcommand{\tnsO}{\tns{0}} 
\newcommand{\tnssI}{\tnss{I}} 
\newcommand{\tnsST}[1]{\boldsymbol{\mathsf{#1}}}

\newcommand{\vv}[1]{\boldsymbol{\mathsf{#1}}}
\newcommand{\mat}[1]{\boldsymbol{\mathsf{#1}}}

\newcommand{\vvO}{\vv{0}}
\newcommand{\matO}{\mat{0}}

\newcommand{\norm}[1]{\left\|#1\right\|}
\newcommand{\normalize}[1]{\frac{#1}{\norm{#1}}}

\newcommand{\relErrorDisp}{\norm{e}_{L2}^{\text{rel}}}

\newcommand{\tr}{^{\mathrm T}}
\newcommand{\trinv}{^{-\mathrm T}}
\newcommand{\inv}{^{-1}}
\newcommand{\dyad}{\otimes}

\renewcommand{\vector}[1]{\begin{bmatrix}#1\end{bmatrix}}
\renewcommand{\matrix}[1]{\begin{bmatrix}#1\end{bmatrix}}

\newcommand{\brackets}[4][]{%
\ifthenelse{\isempty{#1}}{%
\left#2#4\right#3%
}{%
\ifnumcomp{#1}{=}{0}{#2#4#3}
{\ifnumcomp{#1}{=}{1}{\bigl#2#4\bigr#3}
{\ifnumcomp{#1}{=}{2}{\Bigl#2#4\Bigr#3}
{\ifnumcomp{#1}{=}{3}{\biggl#2#4\biggr#3}
{\ifnumcomp{#1}{=}{4}{\Biggl#2#4\Biggr#3}
{size not supported}
}}}}}}

\newcommand{\br}[2][]{\brackets[#1]{(}{)}{#2}}
\newcommand{\brr}[2][]{\brackets[#1]{[}{]}{#2}}
\newcommand{\brrr}[2][]{\brackets[#1]{\{}{\}}{#2}}
\newcommand{\at}[2][]{\brackets[#1]{.}{|}{#2}}

\newcommand{\encapsulate}[1]{{#1}}

\newcommand{\placeholder}{(\cdot)}
\newcommand{\placeholderStar}{(*)}

\DeclareMathOperator{\nl}{nl}

\newcommand{\letterbeamcenterline}{r}
\newcommand{\letterbeampos}{\letterbeamcenterline}
\newcommand{\letterbeamrot}{\theta}

\newcommand{\ebeamName}{e}
\newcommand{\ebeam}{{\elementindex{\ebeamName}}}

\newcommand{\sbeam}{s}

\newcommand{\Ebeam}{E}

\newcommand{\nubeam}{\nu}

\newcommand{\beamlength}{L}

\newcommand{\nbeamn}{n_n}
\newcommand{\nbeamne}{n_n^\ebeam}
\newcommand{\nbeame}{n_{\text{el}}}

\newcommand{\indexbeam}{i}

\newcommand{\sumbeamne}[1]{\sum_{\indexbeam=1}^{\nbeamne}{#1}}

\newcommand{\elementindex}[1]{(#1)}

\newcommand{\nameinternal}{\text{int}}
\newcommand{\nameexternal}{\text{ext}}

\newcommand{\rotMatArbitrary}{\triad^{*}}

\DeclareMathOperator{\rv}{rv}

\newcommand{\triad}{\tnss{\Lambda}}
\newcommand{\triadO}{\tnss{\Lambda}_{0}}

\newcommand{\gtriadO}[1]{\tns{g}_{#1,0}}
\newcommand{\gtriad}[1]{\tns{g}_{#1}}
\newcommand{\dgtriad}[1]{\delta\tns{g}_{#1}}

\newcommand{\rotvec}{\tns{\psi}}
\newcommand{\rotvecArbitrary}{\tns{\psi}^{*}}
\newcommand{\rotvecaxis}{\tns{e}_{\psi}}
\newcommand{\rotvecnorm}{\psi}
\newcommand{\drotvec}{\delta\rotvec}

\newcommand{\rotmult}{\tns{\theta}}
\newcommand{\drotmult}{\delta\rotmult}

\newcommand{\drotmulth}{\delta\rotmult_h}
\newcommand{\drotmulthe}{\delta\rotmult^\ebeam_h}
\newcommand{\drotmultne}{\delta\hat{\rotmult}{}^\ebeam_{\indexbeam}}
\newcommand{\Drotmultn}{\Delta\hat{\rotmult}}

\newcommand{\SskewOp}{\tnss{S}}
\newcommand{\Sskew}[1]{\SskewOp{\br{#1}}}
\newcommand{\Ttrans}{\tnss{T}}
\newcommand{\Qbeamrot}{\vv{d}_{\psi}}
\newcommand{\QbeamPlace}{\vv{d}_{\placeholder}}
\newcommand{\QbeamPlaceStar}{\vv{d}_{\placeholderStar}}
\newcommand{\dQbeamrot}{\delta\vv{d}_{\theta}}
\newcommand{\DQbeamrot}{\Delta\vv{d}_{\psi}}
\newcommand{\DQbeamrotMult}{\Delta\vv{d}_{\theta}}
\newcommand{\DQbeamPlace}{\Delta\vv{d}_{\placeholder}}

\newcommand{\Qbeampos}{\vv{d}_{r}}
\newcommand{\DQbeampos}{\Delta\Qbeampos}
\newcommand{\QbeamVel}{\vv{d}_{v}}
\newcommand{\DQbeamVel}{\Delta\QbeamVel}
\newcommand{\QbeamAngularVel}{\vv{d}_{W}}
\newcommand{\DQbeamAngularVel}{\Delta\QbeamAngularVel}
\newcommand{\dQbeampos}{\delta\Qbeampos}
\newcommand{\dQbeamVel}{\delta\QbeamVel}
\newcommand{\dQbeamAngularVel}{\delta\QbeamAngularVel}

\newcommand{\rbeamn}{\hat{\rbeam}_\indexbeam}
\newcommand{\rbeamni}{\hat{\rbeam}}
\newcommand{\rbeamne}{\hat{\rbeam}^\ebeam_\indexbeam}
\newcommand{\rbeamOn}{\hat{\rbeam}_{0,\indexbeam}}
\newcommand{\rbeamOne}{\hat{\rbeam}^\ebeam_{0,\indexbeam}}

\newcommand{\tbeamn}{\hat{t}_\indexbeam}

\newcommand{\rotvecbeamOn}{\hat{\tns{\psi}}{}_{0,\indexbeam}}

\newcommand{\xisolid}{\xi}
\newcommand{\etasolid}{\eta}

\newcommand{\xisolidDependency}{\br{\xisolid, \etasolid}}

\newcommand{\xisurface}{\theta^1}
\newcommand{\etasurface}{\theta^2}

\newcommand{\indexSurface}{\alpha}
\newcommand{\indexSurfaceAlt}{\beta}
\newcommand{\xisurfaceEinstein}{\theta^\indexSurface}
\newcommand{\tnsCo}[2]{#1_{#2}}
\newcommand{\tnsCont}[2]{#1^{#2}}

\newcommand{\rbeamO}{\tns{r}_0}
\newcommand{\rbeam}{\tns{r}}

\newcommand{\drbeam}{\delta\rbeam}
\newcommand{\drbeamh}{\delta\rbeam_h}

\newcommand{\ubeam}[1][]{\tns{u}_{#1}}

\newcommand{\rbeamOhe}{\tns{r}_{0,h}^\ebeam}

\newcommand{\Nbeamn}{N^\ebeam_\indexbeam}

\newcommand{\Kbeam}{\mat{K}}

\newcommand{\Rbeam}{\vv{r}}
\newcommand{\RbeamPos}{\vv{r}_{\letterbeampos}}
\newcommand{\RbeamRot}{\vv{r}_{\letterbeamrot}}
\newcommand{\RbeamVel}{\vv{r}_{v}}
\newcommand{\RbeamAngularVel}{\vv{r}_{W}}
\newcommand{\RbeamPlace}{\vv{r}_{\placeholder}}

\newcommand{\QbeamPosPos}{\mat{K}_{rr}}
\newcommand{\QbeamPosRot}{\mat{K}_{r\psi}}
\newcommand{\QbeamPosVel}{\mat{K}_{rv}}

\newcommand{\QbeamRotPos}{\mat{K}_{\theta r}}
\newcommand{\QbeamRotRot}{\mat{K}_{\theta\psi}}

\newcommand{\QbeamRotAngularVel}{\mat{K}_{\theta W}}

\newcommand{\QbeamVelPos}{\mat{K}_{vr}}
\newcommand{\QbeamVelVel}{\mat{K}_{vv}}

\newcommand{\QbeamAngularVelRot}{\mat{K}_{W\psi}}
\newcommand{\QbeamAngularVelAngularVel}{\mat{K}_{WW}}

\newcommand{\srTensionMat}{\tns{\Gamma}}
\newcommand{\srTension}{\tns{\gamma}}
\newcommand{\srTensionC}{\tnss{c}_F}
\newcommand{\srTensionCMat}{\tnss{C}_F}
\newcommand{\srBending}{\tns{\omega}}
\newcommand{\srBendingMat}{\tns{\Omega}}
\newcommand{\srBendingC}{\tnss{c}_M}
\newcommand{\srBendingCMat}{\tnss{C}_M}

\newcommand{\rbeamSTO}{\tnsST{r}_0}

\newcommand{\timeVar}{t}

\newcommand{\timehe}{t_h^\ebeam}
\newcommand{\timene}{\hat{t}_\indexbeam^\ebeam}
\newcommand{\timeEnd}{T}
 
\newcommand{\sbeama}{s_a(t)}
\newcommand{\sbeamb}{s_b(t)}
\newcommand{\domainArcLength}{\Omega_l}
\newcommand{\domainArcLengthi}{\Omega^{j}_{l}}
\newcommand{\domainArcLengthihFun}[1]{\Omega^{#1}_{l,h}}
\newcommand{\domainArcLengthih}{\domainArcLengthihFun{j}}
\newcommand{\domainTime}{\mathcal{T}}
\newcommand{\domainSpaceTimeO}{\mathcal{B}_0}
\newcommand{\domainSpaceTimeOh}{\mathcal{B}_{0,h}}
\newcommand{\domainSpaceTime}{\mathcal{B}}

\newcommand{\boundarySpaceA}{\Gamma_{s_a}}
\newcommand{\boundarySpaceB}{\Gamma_{s_b}}
\newcommand{\boundaryTimeO}{\Gamma_{t_0}}
\newcommand{\boundaryTimeT}{\Gamma_{\timeEnd}}

\newcommand{\intSpace}[1]{\int_{\domainArcLength}{ #1 \,\mathrm d \sbeam}}
\newcommand{\intTime}[1]{\int_{\domainTime}{ #1 \,\mathrm d \timeVar}}
\newcommand{\intSpaceTime}[1]{\int_{\domainTime}{ \int_{\domainArcLength(t)}{ #1 \,\mathrm d \sbeam \,\mathrm d \timeVar}}}
\newcommand{\intSpaceTimeFixedMaterial}[1]{\int_{\domainTime}{ \int_{\domainArcLength}{ #1 \,\mathrm d \sbeam \,\mathrm d \timeVar}}}

\newcommand{\angularVelocity}{\tns{w}}
\newcommand{\angularVelocityMat}{\tns{W}}
\newcommand{\angularVelocityMatFromRot}{\tns{W}^{\psi}}
\newcommand{\dangularVelocityMat}{\delta\tns{W}}

\newcommand{\vBeam}{\tns{v}}
\newcommand{\dvBeam}{\delta\tns{v}}
\newcommand{\vBeamFromPos}{\tns{v}^{r}}

\newcommand{\directionSpace}{\tnsST{s}}

\newcommand{\density}{\rho}
\newcommand{\densityRotationMat}{\tnss{C}_{\density}}
\newcommand{\densityRotation}{\tnss{c}_{\density}}
\newcommand{\crossSectionArea}{A}

\newcommand{\beamForce}{\tns{f}}
\newcommand{\beamBodyLoad}{\tilde{\tns{f}}}
\newcommand{\beamForceIntertia}{\tns{f}_{\density}}

\newcommand{\beamMoment}{\tns{m}}
\newcommand{\beamBodyMoment}{\tilde{\tns{m}}}
\newcommand{\beamMomentIntertia}{\tns{m}_{\density}}

\newcommand{\beamLinearMomentum}{\tns{l}}
\newcommand{\beamAngularMomentum}{\tns{h}}

\newcommand{\dW}{\delta W}

\newcommand{\Act}{S}
\newcommand{\dAct}{\delta \Act}
\newcommand{\dActMixed}{\delta \mathcal S}

\newcommand{\stabil}{\theta}

\newcommand{\dvBeamh}{\delta\vBeam_h}

\newcommand{\angularVelocityMathe}{\angularVelocityMat_h^\ebeam}

\newcommand{\dangularVelocityMath}{\delta\angularVelocityMat_h}

\newcommand{\triadRefe}{\triad_{m}^\ebeam}
\newcommand{\triadOffsete}{\triad^{\ebeam}_{i,\text{off}}}

\newcommand{\triadne}{\triad^\ebeam_{\indexbeam}}
\newcommand{\triadOne}{\triad^\ebeam_{\indexbeam,0}}

\newcommand{\triadhe}{\encapsulate{\triad_{h}^\ebeam}}
\newcommand{\triadOhe}{\encapsulate{\triad_{0,h}^\ebeam}}

\newcommand{\rotvecn}{\hat{\rotvec}{}_{\indexbeam}}
\newcommand{\rotvecnO}{\hat{\rotvec}{}_{\indexbeam,0}}
\newcommand{\rotvecneO}{\hat{\rotvec}{}^\ebeam_{\indexbeam,0}}

\newcommand{\Drotvecni}{\Delta\hat{\rotvec}}
\newcommand{\rotvecne}{\hat{\rotvec}{}^\ebeam_{\indexbeam}}

\newcommand{\rotvecneRel}{\tns{\Phi}^\ebeam_{l,\indexbeam}}
\newcommand{\rotvechRel}{\tns{\Phi}^\ebeam_{l,h}}
\newcommand{\rotvechReldt}{\dot{\tns{\Phi}}{}^\ebeam_{l,h}}

\newcommand{\q}{\tns{q}}
\newcommand{\qh}{\q_h}
\newcommand{\qhe}{\q_h^\ebeam}

\newcommand{\qne}{\hat{\q}^\ebeam_{\indexbeam}}
\newcommand{\dqq}{\delta\q}

\newcommand{\dqhe}{\delta\qh^\ebeam}

\newcommand{\dqne}{\delta\hat{\q}^\ebeam_{\indexbeam}}

\newcommand{\beamForceExt}{\tns{f}_\sigma}
\newcommand{\beamMomentExt}{\tns{m}_\sigma}
\newcommand{\domainBoundary}{\Gamma}
\newcommand{\beamCurvature}{\tns{k}}
\newcommand{\beamCurvatureMat}{\tns{K}}
\newcommand{\beamCurvatureMathe}{\beamCurvatureMat^\ebeam_{h}}
\newcommand{\beamCurvatureOMathe}{\beamCurvatureMat^\ebeam_{0,h}}
\newcommand{\dWint}{\dW_{\nameinternal}}
\newcommand{\dWext}{\dW_{\nameexternal}}
\newcommand{\dWInertia}{\dW_{\density}}

%% file: figures/examples_convergence_qualitative_displacement.tex
\tikzsetnextfilename{examples_convergence_qualitative_displacementa}
\begin{tikzpicture}
\begin{groupplot}[
    group style={
        group size=3 by 1,
        horizontal sep=1cm,
    },
    width=0.33\textwidth,
    height=0.27\textwidth,
    legend cell align={left},
    every axis plot/.append style={
        thick,
        mark=none
    },
]

\nextgroupplot[
    title={$u_1$},
    ylabel=$u/\unit{m}$,
    legend to name=grouplegend,
    legend columns=2,
    legend style={
        column sep=8pt,
    },
]
\addplot[solid,blue] table [col sep=comma, x=time, y=u]
    {figures/examples_convergence_qualitative_displacement_AceFEM.csv};
\addlegendentry{ST-\quadFE{4}}

\addplot[dashed,green] table [col sep=comma, x=time, y=u]
    {figures/examples_convergence_qualitative_displacement_4C.csv};
\addlegendentry{\cjShortLine{2}}

\nextgroupplot[
    title={$u_2$},
    xlabel=$t/\unit{s}$,
]
\addplot[solid,blue] table [col sep=comma, x=time, y=v]
    {figures/examples_convergence_qualitative_displacement_AceFEM.csv};

\addplot[dashed,green] table [col sep=comma, x=time, y=v]
    {figures/examples_convergence_qualitative_displacement_4C.csv};

\nextgroupplot[
    title={$u_3$},
]
\addplot[solid,blue] table [col sep=comma, x=time, y=w]
    {figures/examples_convergence_qualitative_displacement_AceFEM.csv};

\addplot[dashed,green] table [col sep=comma, x=time, y=w]
    {figures/examples_convergence_qualitative_displacement_4C.csv};

\end{groupplot}

\node at ($(group c1r1.south)!0.5!(group c3r1.south)$) [below=1.5cm] {\pgfplotslegendfromname{grouplegend}};

\end{tikzpicture}

%% file: figures/examples_convergence_qualitative_energy.tex
\tikzsetnextfilename{examples_convergence_qualitative_energy}
\begin{tikzpicture}
\begin{groupplot}[
    group style={
        group size=3 by 1,
        horizontal sep=1cm,
    },
    width=0.33\textwidth,
    height=0.27\textwidth,
    legend cell align={left},
    every axis plot/.append style={
        thick,
        mark=none
    },
]

\nextgroupplot[
    title={$\Pi_\text{int}$},
    ylabel=$\Pi/\unit{J}$,
    legend to name=grouplegend,
    legend columns=2,
    legend style={
        column sep=8pt,
    },
]
\addplot[solid,blue] table [col sep=comma, x=time, y=internal]
    {figures/examples_convergence_qualitative_energy_AceFEM.csv};
\addlegendentry{ST-\quadFE{4}}

\addplot[dashed,green] table [col sep=comma, x=time, y=internal]
    {figures/examples_convergence_qualitative_energy_4C.csv};
\addlegendentry{\cjShortLine{2}}

\nextgroupplot[
    title={$\Pi_\text{kin}$},
    xlabel=$t/\unit{s}$,
]
\addplot[solid,blue] table [col sep=comma, x=time, y=kinetic]
    {figures/examples_convergence_qualitative_energy_AceFEM.csv};

\addplot[dashed,green] table [col sep=comma, x=time, y=kinetic]
    {figures/examples_convergence_qualitative_energy_4C.csv};

\nextgroupplot[
    title={$\Pi_\text{tot}$},
]
\addplot[solid,blue] table [col sep=comma, x=time, y=total]
    {figures/examples_convergence_qualitative_energy_AceFEM.csv};

\addplot[dashed,green] table [col sep=comma, x=time, y=total]
    {figures/examples_convergence_qualitative_energy_4C.csv};

\end{groupplot}

\node at ($(group c1r1.south)!0.5!(group c3r1.south)$) [below=1.5cm] {\pgfplotslegendfromname{grouplegend}};

\end{tikzpicture}

%% file: figures/examples_convergence_uniform_quad4.tex
\tikzsetnextfilename{examples_convergence_uniform_quad4}
\begin{tikzpicture}

\begin{axis}[
    title=ST-\quadFE{4},
    xmode=log,
    ymode=log,
    scale only axis,
    xlabel=$r$,
    ylabel=$\relErrorDisp$,
    legend pos=outer north east,
    legend cell align={left},
    log basis x={2},
    legend to name=examples_convergence_uniform_legend,
    legend columns=4,
    legend style={
        column sep=8pt,
    },
    ]

\addplot+ [] table [col sep=comma, x expr={0.1/\thisrow{time_step}}, y=quad4_0.0] {figures/examples_convergence_uniform.csv};
\addlegendentry{$\stabil=0$}
\addplot+ [] table [col sep=comma, x expr={0.1/\thisrow{time_step}}, y=quad4_0.0001] {figures/examples_convergence_uniform.csv};
\addlegendentry{$\stabil=0.0001$}
\addplot+ [] table [col sep=comma, x expr={0.1/\thisrow{time_step}}, y=quad4_0.001] {figures/examples_convergence_uniform.csv};
\addlegendentry{$\stabil=0.001$}

\addplot[
    domain=4:64,
    samples=2,
    dashed
] {5*x^(-2)};
\addlegendentry{$\order{r^{-2}}$}

\addplot+ [] table [col sep=comma, x expr={0.1/\thisrow{time_step}}, y=quad4_0.01] {figures/examples_convergence_uniform.csv};
\addlegendentry{$\stabil=0.01$}
\addplot+ [] table [col sep=comma, x expr={0.1/\thisrow{time_step}}, y=quad4_0.1] {figures/examples_convergence_uniform.csv};
\addlegendentry{$\stabil=0.1$}
\addplot+ [] table [col sep=comma, x expr={0.1/\thisrow{time_step}}, y=quad4_0.3] {figures/examples_convergence_uniform.csv};
\addlegendentry{$\stabil=0.3$}

\addlegendimage{black,dashdotted}
\addlegendentry{$\order{r^{-3}}$}

\end{axis}
\end{tikzpicture}

%% file: figures/examples_convergence_uniform_quad9.tex
\tikzsetnextfilename{examples_convergence_uniform_quad9}
\begin{tikzpicture}
\begin{axis}[
    title=ST-\quadFE{9},
    xmode=log,
    ymode=log,
    scale only axis,
    xlabel=$r$,
    ylabel=$\relErrorDisp$,
    legend pos=outer north east,
    legend cell align={left},
    log basis x={2},
    ]
\addplot+ [] table [col sep=comma, x expr={0.1/\thisrow{time_step}}, y=quad9_0.0] {figures/examples_convergence_uniform.csv};
\addplot+ [] table [col sep=comma, x expr={0.1/\thisrow{time_step}}, y=quad9_0.0001] {figures/examples_convergence_uniform.csv};
\addplot+ [] table [col sep=comma, x expr={0.1/\thisrow{time_step}}, y=quad9_0.001] {figures/examples_convergence_uniform.csv};

\addplot[
    domain=4:32,
    samples=2,
    dashdotted
] {0.25*x^(-3)};

\addplot+ [] table [col sep=comma, x expr={0.1/\thisrow{time_step}}, y=quad9_0.01] {figures/examples_convergence_uniform.csv};
\addplot+ [] table [col sep=comma, x expr={0.1/\thisrow{time_step}}, y=quad9_0.1] {figures/examples_convergence_uniform.csv};
\addplot+ [] table [col sep=comma, x expr={0.1/\thisrow{time_step}}, y=quad9_0.3] {figures/examples_convergence_uniform.csv};

\end{axis}
\end{tikzpicture}

%% file: figures/examples_convergence_uniform_all.tex
\tikzsetnextfilename{examples_convergence_uniform_all}
\begin{tikzpicture}
\begin{axis}[
    xmode=log,
    ymode=log,
    scale only axis,
    xlabel=$r$,
    ylabel=$\relErrorDisp$,
    legend pos=outer north east,
    legend cell align={left},
    log basis x={2},
    ]

\addplot+ [] table [col sep=comma, x expr={0.1/\thisrow{time_step}}, y=quad4_0.1] {figures/examples_convergence_uniform.csv};
\addlegendentry{ST-\quadFE{4}, $\stabil=0.1$}

\addplot+ [] table [col sep=comma, x expr={0.1/\thisrow{time_step}}, y=quad9_0.1] {figures/examples_convergence_uniform.csv};
\addlegendentry{ST-\quadFE{9}, $\stabil=0.1$}

\addplot+ [] table [col sep=comma, x expr={0.1/\thisrow{time_step}}, y=line2] {figures/examples_convergence_uniform.csv};
\addlegendentry{\cjShortLine{2}}
\addplot+ [] table [col sep=comma, x expr={0.1/\thisrow{time_step}}, y=line3] {figures/examples_convergence_uniform.csv};
\addlegendentry{\cjShortLine{3}}

\addplot[
    domain=4:64,
    samples=2,
    dashed
] {60*x^(-2)};
\addlegendentry{$\order{r^{-2}}$}
\addplot[
    domain=4:32,
    samples=2,
    dashdotted
] {0.25*x^(-3)};
\addlegendentry{$\order{r^{-3}}$}

\end{axis}
\end{tikzpicture}

%% file: figures/examples_convergence_stabilization.tex
\tikzsetnextfilename{examples_convergence_stabilization_normal}
\begin{tikzpicture}
\begin{axis}[
    scale only axis,
    xlabel=$t/\unit{s}$,
    ylabel=$u_3/\unit{m}$,
    legend pos=outer north east,
    legend cell align={left},
    legend to name=examples_convergence_stabilization_legend,
    legend columns=7,
    legend style={
        column sep=8pt,
    },
    ]

\addplot+ [] table [col sep=comma, x=time, y=0.0] {figures/examples_convergence_stabilization.csv};
\addlegendentry{$\stabil=0$}

\addplot+ [] table [col sep=comma, x=time, y=0.0001] {figures/examples_convergence_stabilization.csv};
\addlegendentry{$\stabil=0.0001$}

\addplot+ [] table [col sep=comma, x=time, y=0.001] {figures/examples_convergence_stabilization.csv};
\addlegendentry{$\stabil=0.001$}

\addplot+ [] table [col sep=comma, x=time, y=0.01] {figures/examples_convergence_stabilization.csv};
\addlegendentry{$\stabil=0.01$}

\addplot+ [] table [col sep=comma, x=time, y=0.1] {figures/examples_convergence_stabilization.csv};
\addlegendentry{$\stabil=0.1$}

\addplot+ [] table [col sep=comma, x=time, y=0.3] {figures/examples_convergence_stabilization.csv};
\addlegendentry{$\stabil=0.3$}

\end{axis}
\end{tikzpicture}
\hfill
\tikzsetnextfilename{examples_convergence_stabilization_log}
\begin{tikzpicture}
\begin{axis}[
    ymode=log,
    scale only axis,
    xlabel=$t/\unit{s}$,
    ylabel=$\norm{u_3}/\unit{m}$,
    legend pos=outer north east,
    legend cell align={left},
    legend to name=dummy,
    ]

\addplot+ [] table [col sep=comma, x=time, y expr={abs(\thisrow{0.0})}] {figures/examples_convergence_stabilization.csv};
\addlegendentry{$\stabil=0$}

\addplot+ [] table [col sep=comma, x=time, y expr={abs(\thisrow{0.0001})}] {figures/examples_convergence_stabilization.csv};
\addlegendentry{$\stabil=0.0001$}

\addplot+ [] table [col sep=comma, x=time, y expr={abs(\thisrow{0.001})}] {figures/examples_convergence_stabilization.csv};
\addlegendentry{$\stabil=0.001$}

\addplot+ [] table [col sep=comma, x=time, y expr={abs(\thisrow{0.01})}] {figures/examples_convergence_stabilization.csv};
\addlegendentry{$\stabil=0.01$}

\addplot+ [] table [col sep=comma, x=time, y expr={abs(\thisrow{0.1})}] {figures/examples_convergence_stabilization.csv};
\addlegendentry{$\stabil=0.1$}

\addplot+ [] table [col sep=comma, x=time, y expr={abs(\thisrow{0.3})}] {figures/examples_convergence_stabilization.csv};
\addlegendentry{$\stabil=0.3$}

\end{axis}
\end{tikzpicture}
\tikzset{external/export next=false}
\tikzsetnextfilename{examples_convergence_stabilization_legend}
\pgfplotslegendfromname{examples_convergence_stabilization_legend}

%% file: figures/examples_free_floating_l_deformed.tex
\begin{tikzpicture}
\node[anchor=south west,inner sep=-0.2pt] (image) at (0,0) {\includegraphics[scale=0.24]{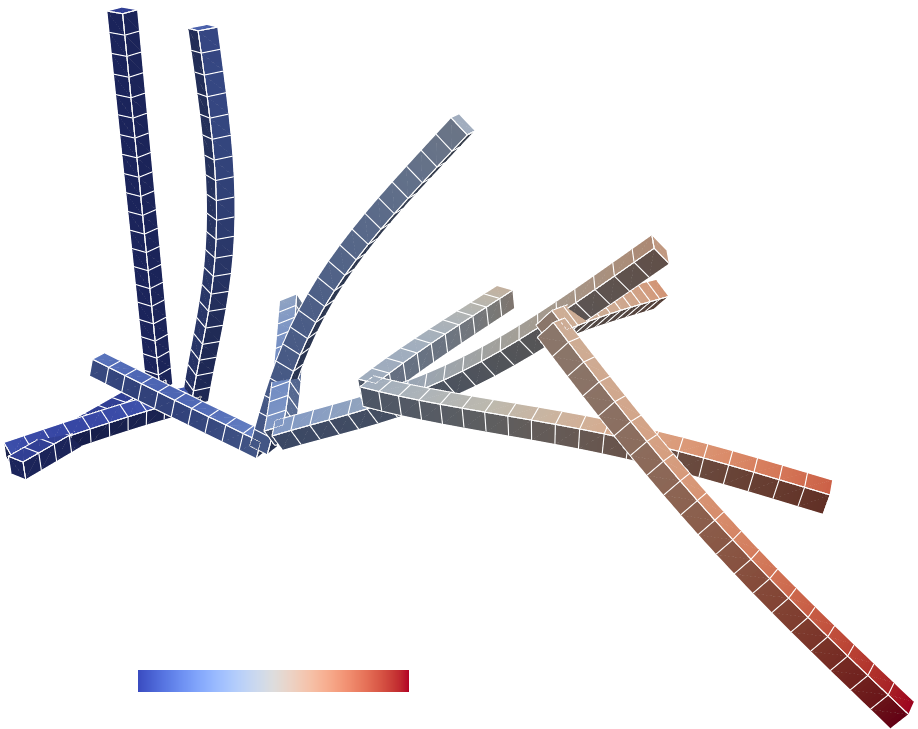}};
\begin{axis}[
scale only axis,
scaled x ticks=false,
scaled y ticks=false,
at={(1.1684cm,0.3132666666666667cm)},
tick label style={font=\footnotesize},
title=$\norm{\tns{u}}/\unit{m}$,
xticklabel={$\pgfmathprintnumber[sci,precision=1,sci generic={mantissa sep=,exponent={\mathrm{e}{##1}}}]{\tick}$},
yticklabel={$\pgfmathprintnumber[sci,precision=1,sci generic={mantissa sep=,exponent={\mathrm{e}{##1}}}]{\tick}$},
xmin=1.4630045988600737e-14,
xmax=4.597403838616087,
ymin=1.4630045988600737e-14,
ymax=4.597403838616087,
width=2.2944666666666667cm,
height=0.18626666666666666cm,
ytick=\empty,
xtick={1.4630045988600737e-14,4.597403838616087},
xtick pos=right,
xtick align=outside,
title style={{yshift=10pt,}},
]
\end{axis}

\begin{scope}[x={(image.south east)},y={(image.north west)}]


    \newcommand{\annotate}[7]{%
    \coordinate (P) at (#1,#2);
    \draw ($(P)+(#3:#5)$) -- ($(P)+(#3:#4)$)
        node[inner sep=1pt, anchor=#6] {\footnotesize$t=\unit[#7]{s}$};
    }

    \annotate{0.13}{0.975}{90}{0.75cm}{3mm}{south}{0}
    \annotate{0.22}{0.96}{30}{1cm}{3mm}{south west}{2}
    \annotate{0.5}{0.825}{30}{1cm}{3mm}{south west}{4}
    \annotate{0.71}{0.65}{30}{1cm}{3mm}{south west}{6}
    \annotate{0.9}{0.325}{60}{1cm}{3mm}{south}{8}
    \annotate{0.975}{0.025}{90}{1cm}{5mm}{south}{10}

\end{scope}

\end{tikzpicture}%

%% file: figures/examples_free_floating_l_displacement.tex
\tikzsetnextfilename{examples_free_floating_l_displacement}
\begin{tikzpicture}
\begin{groupplot}[
    group style={
        group size=3 by 1,
        horizontal sep=1cm,
    },
    width=0.33\textwidth,
    height=0.27\textwidth,
    legend cell align={left},
    every axis plot/.append style={
        thick,
        mark=none
    },
]

\nextgroupplot[
    title={$u_1$},
    ylabel=$u/\unit{m}$,
    legend to name=grouplegend,
    legend columns=2,
    legend style={
        column sep=8pt,
    },
]
\addplot[solid,blue] table [col sep=comma, x=time, y=x]
    {figures/examples_free_floating_l_displacement_ace_fem.csv};
\addlegendentry{ST-\quadFE{9}}

\addplot[dashed,green] table [col sep=comma, x=time, y=x]
    {figures/examples_free_floating_l_displacement_four_c.csv};
\addlegendentry{\cjShortLine{3}}

\nextgroupplot[
    title={$u_2$},
    xlabel=$t/\unit{s}$,
]
\addplot[solid,blue] table [col sep=comma, x=time, y=z]
    {figures/examples_free_floating_l_displacement_ace_fem.csv};

\addplot[dashed,green] table [col sep=comma, x=time, y=z]
    {figures/examples_free_floating_l_displacement_four_c.csv};

\nextgroupplot[
    title={$u_3$},
]
\addplot[solid,blue] table [col sep=comma, x=time, y=y]
    {figures/examples_free_floating_l_displacement_ace_fem.csv};

\addplot[dashed,green] table [col sep=comma, x=time, y=y]
    {figures/examples_free_floating_l_displacement_four_c.csv};

\end{groupplot}

\node at ($(group c1r1.south)!0.5!(group c3r1.south)$) [below=1.5cm] {\pgfplotslegendfromname{grouplegend}};

\end{tikzpicture}

%% file: figures/examples_sliding_beam_mesh.tex
\tikzsetnextfilename{examples_sliding_beam_mesh}
\begin{tikzpicture}[
    x=0.95cm,
    y=0.95cm,
    line cap=round,
    line join=round
]


\def\BaseH{5}
\def\BaseR{5}

\def\Size{4}

\pgfmathsetmacro{\GeomScale}{\Size/\BaseH}

\pgfmathsetmacro{\H}{\GeomScale*\BaseH}
\pgfmathsetmacro{\R}{\GeomScale*\BaseR}

\pgfmathdeclarefunction{sliding}{1}{%
    \pgfmathparse{%
        (#1 <= 1)
        ? pow(#1,3)/6
        : (
            (#1 <= 2)
            ? 1/3 - #1 + pow(#1,2) - pow(#1,3)/6
            : -1 + #1
          )
    }%
}

%
\pgfmathdeclarefunction{L}{1}{%
    \pgfmathparse{%
        \GeomScale*sliding(\BaseH*#1)
    }%
}

\pgfmathdeclarefunction{X}{2}{%
    \pgfmathparse{%
        (1-#1)*L(#2) + #1*\R
    }%
}

\pgfmathdeclarefunction{Y}{2}{%
    \pgfmathparse{%
        \H*#2
    }%
}

\path[fill=gray!62]
    ({L(0)},0)
    \foreach \k in {1,...,100} {
        -- ({L(\k/100)},{\H*\k/100})
    }
    -- (\R,\H)
    -- (\R,0)
    -- cycle;

\foreach \j in {0,...,20}{
    \pgfmathsetmacro{\v}{\j/20}

    \draw[black!65]
        ({L(\v)},{\H*\v})
        --
        (\R,{\H*\v});
}

\foreach \i in {0,...,15}{
    \pgfmathsetmacro{\u}{\i/15}

    \draw[black!65]
        plot[
            samples=120,
            domain=0:1
        ]
        ({X(\u,\x)},{Y(\u,\x)});
}

\draw[black,thick]
    ({L(0)},0)
    \foreach \k in {1,...,150} {
        -- ({L(\k/150)},{\H*\k/150})
    }
    -- (\R,\H)
    -- (\R,0)
    -- cycle;

\draw[
    -{Latex[length=1.5ex,width=0.75ex]}
]
    (1.025*\R,0)
    --
    (1.2*\R,0)
    node[below] {$\sbeam$};

\draw[
    -{Latex[length=1.5ex,width=0.75ex]}
]
    (0,0.2*\H)
    --
    (0,{1.15*\H})
    node[left] {$t$};

\draw (0,-0.2) -- (0,-0.08) node[below=3pt] {$0$};
\draw (\R,-0.2) -- (\R,-0.08) node[below=3pt] {$L$};

\draw (-0.2,0) -- (-0.08,0) node[left=3pt] {$0$};
\draw (-0.12,\H) -- (-0.0,\H) node[left=3pt] {$\timeEnd$};

\coordinate (a) at ({L(0.5)},{\H*0.5});
\coordinate (b) at (\R,{\H*0.5});
\node[right] at (b) {$l(t)$};
\draw[-{Latex[length=1.5ex,width=0.75ex]}, thick] (a) -- (b);
\fill (a) circle[radius=2pt];

\coordinate (boundA) at ({L(0.7)},{\H*0.7});
\draw[shorten <=3pt, shorten >=-5pt] (boundA) -- ++(-0.5,0.5) node[anchor=south east] {$\boundarySpaceA$};

\coordinate (boundB) at ({\R},{\H*0.7});
\draw[shorten <=3pt, shorten >=-6pt] (boundB) -- ++(0.5,0.5) node[anchor=south west] {$\boundarySpaceB$};

\end{tikzpicture}

%% file: figures/examples_sliding_beam_configuration.tex
\tikzsetnextfilename{examples_sliding_beam_configuration}
\begin{tikzpicture}
\usetikzlibrary{calc}
\begin{axis}[
    scale only axis,
    xlabel=$r_1/\unit{m}$,
    ylabel=$r_2/\unit{m}$,
    legend pos=outer north east,
    legend cell align={left},
    log basis x={2},
    every axis plot/.append style={
        thick,
        mark=none
    },
    xmax=5,
    ymax=1.5,
    ]

\addplot [thin, lightgray] table [col sep=comma, x=x, y=y] {figures/examples_sliding_beam_configuration_all.csv};
\addplot [thick, black] table [col sep=comma, x=x, y=y] {figures/examples_sliding_beam_configuration_bound.csv};

\coordinate (minBox) at (rel axis cs:0,0);
\coordinate (maxBox) at (rel axis cs:1,1);
\coordinate (center) at (axis cs:0,0);
\draw[dashed, gray] (center -| minBox) -- (center -| maxBox);
\draw[dashed, gray] (center |- minBox) -- (center |- maxBox);

\coordinate (P) at (axis cs:4.5034446636166825,-2.0287846283232653);
\draw[{Latex}-,shorten <=3pt] (P) -- ++(0.0cm,0.5cm) node[anchor=south] {\footnotesize$t=\unit[0]{s}$};

\coordinate (P) at (axis cs:2.486213554760864,0.5236652214587579);
\draw[{Latex}-,shorten <=3pt] (P) -- ++(0.6cm,0.2cm) node[anchor=west] {\footnotesize$t=\unit[3.45]{s}$};

\coordinate (P) at (axis cs:1.7394875894473478,-0.5899338724151132);
\draw[{Latex}-,shorten <=3pt] (P) -- ++(0.5cm,-0.7cm) node[anchor=north] {\footnotesize$t=\unit[4.15]{s}$};

\coordinate (P) at (axis cs:1.333403166374123,0.466384453926333);
\draw[{Latex}-,shorten <=3pt,shorten >=-3pt] (P) -- ++(0.5cm,0.4cm) node[anchor=south west] {\footnotesize$t=\unit[4.58]{s}$};

\coordinate (P) at (axis cs:1.0673052411166966,-0.4399702914975071);
\draw[{Latex}-,shorten <=3pt] (P) -- ++(-0.3cm,-0.5cm) node[anchor=north] {\footnotesize$t=\unit[4.83]{s}$};

\coordinate (P) at (axis cs:0.9112082588131161,0.38041942174319976);
\draw[{Latex}-,shorten <=3pt] (P) -- ++(0.0cm,0.5cm) node[anchor=south] {\footnotesize$t=\unit[5]{s}$};

\end{axis}
\end{tikzpicture}

%% file: figures/examples_sliding_beam_tip_displacement.tex
\tikzsetnextfilename{examples_sliding_beam_tip_displacement}
\begin{tikzpicture}

\begin{axis}[
    scale only axis,
    xlabel=$r_1/\unit{m}$,
    ylabel=$r_2/\unit{m}$,
    legend pos=south west,
    legend cell align={left},
    log basis x={2},
    every axis plot/.append style={
        thick,
        mark=none
    },
    ]

\addplot [solid,blue] table [col sep=comma, x=u, y=v] {figures/examples_sliding_beam_tip_displacement_ace_fem.csv};
\addlegendentry{ST-\quadFE{4}}
\addplot [dashed,green] table [col sep=comma, x=u, y=v] {figures/examples_sliding_beam_tip_displacement_sbf.csv};
\addlegendentry{SBF}

\end{axis}
\end{tikzpicture}